\documentclass{article}
\usepackage[utf8]{inputenc}

\usepackage[parfill]{parskip}
\usepackage{caption}
\usepackage{tabularx}
\usepackage{latexsym}
\usepackage{amstext,amsfonts,amssymb,amsmath}
\usepackage{dsfont}
\usepackage{subfig}
\usepackage{todonotes}
\usepackage{authblk}

\usepackage{multicol}
\usepackage{nonfloat}
\usepackage{color}
\usepackage{bm}
\usepackage[margin=1in]{geometry}
\usepackage{graphicx}
\usepackage{wrapfig}
\usepackage{enumitem}
\usepackage[font={small}]{caption}

\title{Experimental Data Based Reduced Order Model for Analysis and Prediction of Flame Transition in Gas Turbine Combustors}
\date{}
\author[1]{Shivam Barwey\thanks{Contact: sbarwey@umich.edu. Preprint accepted to Combustion Theory and Modelling.}}
\author[1]{Malik Hassanaly}
\author[2]{Qiang An}
\author[1]{Venkat Raman}
\author[2]{Adam Steinberg}
\affil[1]{Department of Aerospace Engineering, University of Michigan}
\affil[2]{Institute for Aerospace Studies, University of Toronto}

\begin{document}
\maketitle

\begin{abstract}
In lean premixed combustors, flame stabilization is an important operational concern that can affect efficiency, robustness and pollutant formation. The focus of this paper is on flame lift-off and re-attachment to the nozzle of a swirl combustor. Using time-resolved experimental measurements, a data-driven approach known as cluster-based reduced order modeling (CROM) is employed to 1) isolate key flow patterns and their sequence during the flame transitions, and 2) formulate a forecasting model to predict the flame instability. The flow patterns isolated by the CROM methodology confirm some of the experimental conclusions about the flame transition mechanism. In particular, CROM highlights the key role of the precessing vortex core (PVC) in the flame detachment process in an unsupervised manner. For the attachment process, strong flow recirculation far from the nozzle appears to drive the flame upstream, thus initiating re-attachment. Different data-types (velocity field, OH concentration) were processed by the modeling tool, and the predictive capabilities of these different models are also compared. It was found that the swirling velocity possesses the best predictive properties, which gives a supplemental argument for the role of the PVC in causing the flame transition. The model is tested against unseen data and successfully predicts the probability of flame transition (both detachment and attachment) when trained with swirling velocity with minimal user input. The model trained with OH-PLIF data was only successful at predicting the flame attachment, which implies that different physical mechanisms are present for different types of flame transition. Overall, these aspects show the great potential of data-driven methods, particularly probabilistic forecasting techniques, in analyzing and predicting large-scale features in complex turbulent combustion problems. \\
\textit{Keywords: Reduced order modeling; Data-driven modeling; Flame transition; Probabilistic forecasting; Premixed combustion}
\end{abstract}

\tableofcontents

\section{Introduction}

The issue of flame stabilization is critical for lean premixed flames with the perspective of controlling pollutant formation and ensuring robustness at various operating loads \cite{qiang}. In swirl-stabilized combustors, the interaction of multiple recirculation zones with flame propagation impacts the stabilization process. Two different stabilization mechanisms are feasible \cite{qiang16,qiang17,qiang18}: a) a shear layer stabilized flame resulting in a flame attachment to some feature of the geometry, and b) lifted flames that are stabilized near the stagnation surface formed between inflowing gases and an inner recirculation zone. For practical considerations, the transition into a lifted state might serve as a precursor for flame blow-off \cite{stohr_blowout}. Prior studies have focused on elucidating physical mechanisms that cause transition between the attached and lifted flame states, which are due to either inflow or other operational variations. The focus of this work is to use one such experimental study to develop a prognostic data-driven model with a twofold purpose: 1) to understand how the flame instability occurs and 2) to predict the flame transition.

Instabilities in complex systems such as gas turbines are subject to multi-scale mechanisms that require a simplified representation to be meaningful. A first technique is to quantify the growth rates of perturbations, which are obtained using system identification (SI) methods \cite{noiray}. The perturbations are quantified for some quantity of interest, e.g. pressure at some location. The SI technique can be input-output based, where the response of the combustor to perturbations are used, or output based \cite{polifkeSI,noiray9,noiray}. A second technique utilizes data decomposition to obtain a simplified representation of the governing equations. Proper orthogonal decomposition (POD) \cite{berkoozPOD,steinbergPOD} projects the governing equations onto some meaningful flow modes, and dynamic mode decomposition (DMD) attempts to linearize the governing equations \cite{schmidDMD}. Such methods determine instability modes from experimental data and the variation of their relative contribution to the flow field. These characteristic modes can be used to directly analyze physical system properties of interest, and also to construct a reduced-order model (ROM) of the system. These approaches have been used in the past for combustion chambers  \cite{ananthkrishnan2005,motheauDMD}, and additional studies have utilized laser diagnostics to determine characteristic modes \cite{graftieaux2001, oberleithner_pod}.

From a prognostic viewpoint, the above methods are derived from two different processes. SI and DMD-related techniques are purely data-driven approaches, where finite measurements of the system are used to construct empirical tools which may then be used to predict combustor state transitions. While such tools are easier to use in a practical setting, interpreting the model in terms of the physical processes is not straightforward. Further, DMD related methods invoke linearization of the underlying dynamics, which may not be valid in many practical configurations. In contrast, POD-type techniques attempt to simplify the theoretical governing equations to predict the future states of the system. First, data from the system is used to construct lower-order representations, or a ROM; second, this model can be propagated in time in order to predict the future state of the system. Since the decomposition techniques can be related to physical processes, the first step provides meaningful projection operators which can be used to analyze the physics of the combustion. However, the projection applied to the governing equations leads to issues with closure and numerical discretization, similar to those encountered in reduced-fidelity models such as RANS/LES \cite{pope_10,pope_jfm,ramanEmergingTrends}.

Recently, a so-called cluster-based reduced-order modeling (CROM) framework has been proposed by Kaiser et al.~\cite{kaiser_crom}. Unlike prior decomposition tools that obtain a linear operator to describe the nonlinear dynamics, CROM retains the nonlinearity of the underlying system. Furthermore, CROM avoids a direct projection of the governing equations, thereby alleviating the closure problem; instead, it constructs a discrete-time Markov process from a set of experimental or high-resolution simulation data. As a result, CROM provides a physically-meaningful decomposition of the dataset which allows to understand from the available data how the flame instability occurs~\cite{hassanaly2019computational}. It also provides a probabilistic model for the forecast of the combustor state. In the original work of Kaiser et al.~\cite{kaiser_crom}, CROM was used to construct the model of a turbulent mixing layer. This method has also been used to predict cycle-to-cycle variations in internal combustion engines via cluster discretization of data \cite{cao_crom}. 

In this study, CROM is used to develop a means for transition mechanism analysis and to model flame topology transition in a swirl combustor. Experimental stereoscopic particle image velocimetry (S-PIV) and OH planar laser induced fluorescence (PLIF) images are used to develop a predictive ROM for flame transition. The flow modes obtained are analyzed to illustrate the physical information that can be obtained from the CROM methodology. The remainder of this paper is organized as follows: experimental data and operating conditions are provided in Sec.~\ref{sec:data_used}; the CROM methodology is discussed in Sec.~\ref{sec:crom}; outcomes of the CROM approach are used to gain physical intuition about the flame transition mechanism in Sec.~\ref{sec:results-centroids}; results of the CROM approach both in terms of the prediction horizon time and forecast capability for different data types are discussed in Sec.~\ref{sec:results-forecasting}. Finally, conclusions and future directions are presented in Sec.~\ref{sec:conclusion}.

\section{Experimental Configuration and Dataset}
\label{sec:data_used}
The data used for this analysis was acquired in the experiments of Ref. \cite{qiang_new}, which contains additional details on the combustor and diagnostic techniques. A brief summary is provided here. 

\begin{figure}
    \centering
    \includegraphics[width = 1\columnwidth]{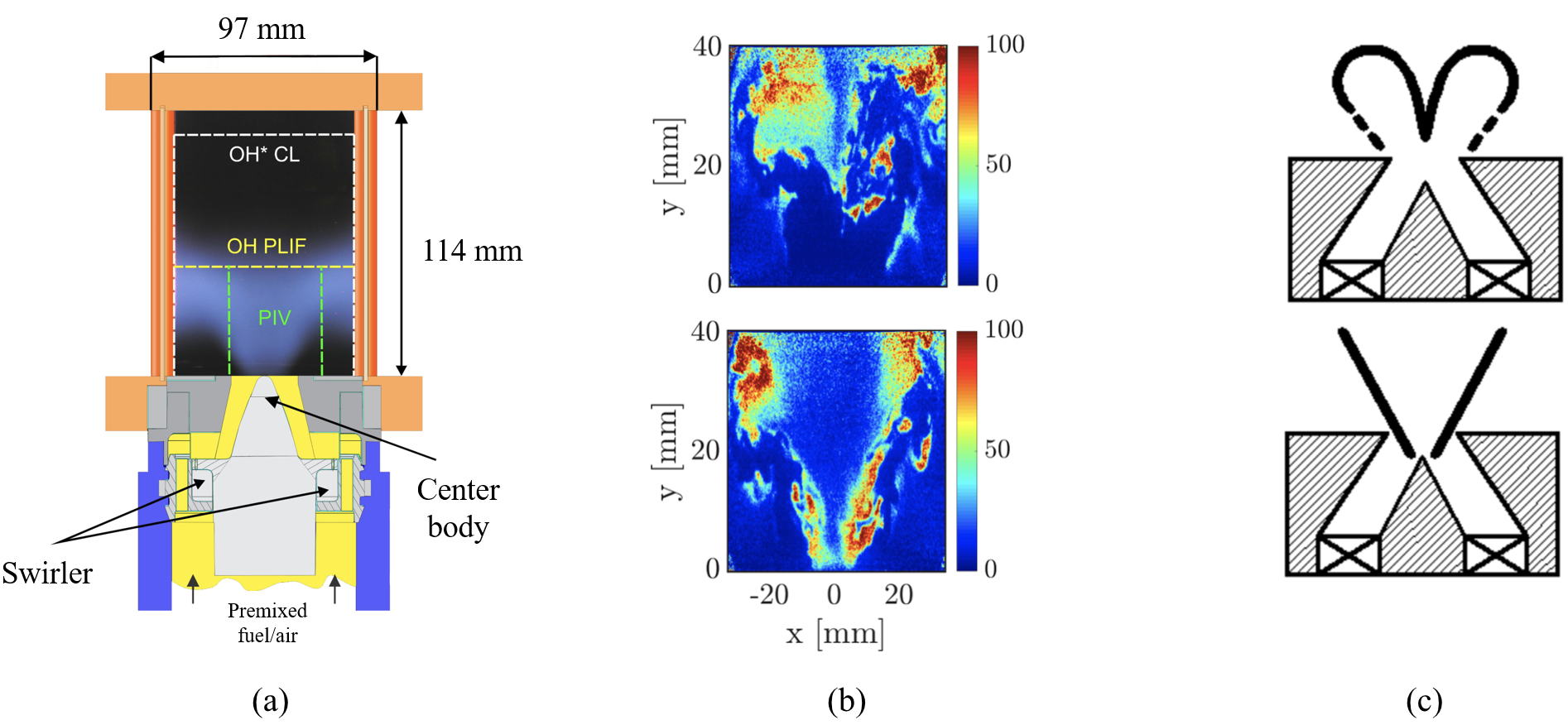}
    \vspace{-15pt}
    \caption{(a) DLR combustor schematic. The fuel consists of $80\%$ $CH_4$ and $20\%$ $CO_2$ by volume. (b) OH-PLIF snapshots of detached (top) and attached (bottom) flames in units of relative pixel intensity. (c) Sketches of M-shaped detached flame (top) and V-shaped attached flame (bottom) (from Ref.~\cite{qiang}).}
    \label{fig:schematic_and_flame}
\end{figure}

Figure~\ref{fig:schematic_and_flame}a shows the gas turbine model combustor, which has been used for a number of previous studies, both experimental \cite{caux-brisebois, meier, oconnor1} and numerical \cite{koo_dlr}. A review addressing progress and key findings in experimental studies on hydrodynamic instabilities, including additional perspective on the advancements in numerical simulations of instabilities in swirling combustors is provided in Ref.~\cite{qiang1}. The schematic shown in Figure~\ref{fig:schematic_and_flame}c depicts the type of flame shape geometry observed in the attached and lifted states as seen by this particular swirl combustor. 

Premixed fuel and air are fed through a plenum to the radial swirler before entering the combustion chamber, where vortex breakdown generates a strong inner recirculation zone. In the present case, a fuel-air mixture of equivalence ratio of $\phi = 0.60$ is fed to the combustor at a preheated temperature of 400~K. The fuel is made of 80\% CH$_4$ and 20\% CO$_2$ by volume. The air flow rate is 400~SLPM. This case was selected from the test matrix in Ref.~\cite{qiang_new} because it exhibits a high number of transitions between clearly defined attached and detached flame states (8 observed in a span of 1.5s), while operating at fixed equivalence ratio and flow rates. Thus, the flow conditions were not changed to force a transition, and the inherent system can be considered ergodic; the flame experienced intermittent and spontaneous (in the sense of being apparently random in time) transitions between the detached and lifted states. 

Data was collected using 10~kHz repetition-rate OH PLIF and S-PIV, providing simultaneous time-resolved 2D measurements of the OH radical distribution and of the three velocity components over a time span of 1.5~s. Figure~\ref{fig:schematic_and_flame}b shows typical instantaneous OH PLIF images in attached and detached states. In the attached state, high OH concentrations are present in the shear layers that separate the inner and outer recirculation zones, while in the detached state, a rotating helical vortex core generates a highly asymmetric OH field. Attached and detached flame configuration shapes are shown in the drawings in Fig.~\ref{fig:schematic_and_flame}c -- the flame in the attached state takes on the characteristic V-shape, whereas the M-shape is observed in the detached. There were roughly 8 total transitions captured in the 1.5~s dataset (combining attached-to-detached transitions and the detached-to-attached transitions). For the specified operating conditions, the flow-through time is 14.6~ms (from Ref.~\cite{qiang_new}) and the flame transition timescale (determined by manually observing the 8 detached-to-attached and attached-to-detached transition times in the OH-PLIF images) is on the order of 10~ms. 

In order to reduce the computational overhead, the PLIF images were reduced in size from 832 $\times$ 504 pixels to 104 $\times$ 63 pixels via nearest-neighbor interpolation based filtering, which preserves the large-scale flame features. The effective resolution of the down-sampled OH PLIF was 0.71~mm. The full-resolution PIV data was used (79 $\times$ 53), with a vector spacing of 0.78~mm and interrogation box size of 1.56~mm. Note that in Fig.~\ref{fig:schematic_and_flame}a, the PIV window is not perfectly symmetric about the combustor centerline, which is reflected accordingly in the figures and discussion below. Nevertheless, the procedure used here only requires the data to coincide in time. Different datasets spanning different parts of the combustors can therefore be used.

\section{Cluster-Based ROM Methodology}

\label{sec:crom}
In this section, the CROM methodology is described. Essentially, it processes time-resolved data of any kind, identifies recurrent patterns in the system dynamics, and creates a probabilistic predictive model for these patterns. Here, the data are the 2D experimental OH-PLIF and PIV images. The set of experimental images is denoted by ${\cal S} = \{S_1, S_2,\cdots,S_N\}$, where $N$ is the number of snapshots available and each subscript denotes a different snapshot. The snapshots are ordered in time. This time-ordered quality is important to create the predictive model, but not important when classifying the data. Each element $S_i$ of the set $\mathcal{S}$ is a vector of pixel values associated with PIV and/or OH-PLIF images, and is of size $N_p$ (i.e, $S_i \in \mathbb{R}^{N_p}$). Snapshots are separated by a time-step $\Delta t$ that is determined by the 10~kHz sampling rate identical across all measurement types. In this section, the data classification (clustering) and the probabilistic model (transition matrix) obtained from the CROM methodology are described. Further, a procedure to relate the cluster outputs to the individual swirler states is also presented (a systematic labeling of the clusters as detached flame, attached flame or transitioning flame).

\subsection{Clustering}
\label{sec:crom:clustering}
The first step of the CROM methodology is to map the set of snapshots ${\cal S}$ to a smaller set of so-called \textit{centroids} ${\cal C} =\{C_1, C_2,\cdots, C_{N_k}\}$, where $C_i \in \mathbb{R}^{N_p}$. The centroids are also images made of $N_p$ pixels that represent some pattern of the flow field. Centroids can be interpreted as delta functions that discretize the probability density function (PDF) of the states. States nearby one another are represented by the same delta function. This step is reminiscent of classification methods used in language interpretation: the same word can be pronounced differently, but a language interpreter finds common characteristics of these different signals to assign them the same meaning. The number of delta functions (centroids) $N_k$ chosen to discretize the state-space is defined by the user, varying within the bounds of 1 to $N$ ($N$ being the number of snapshots). The value of the parameter $N_k$ should be set depending on the purpose of the study, as discussed in Sec.~\ref{sec:num_cluster_transition} and Sec.~\ref{sec:number_of_clusters_prediction}. Each centroid $C_i$ represents a region of phase space that contains some proportion of the total number of snapshots. This region of phase space, denoted $\mathbb{C}_i$, is called a \textit{cluster}. Since there is one centroid per cluster, $N_k$ defines both the number of centroids and the number of clusters.

The $N_k$ centroids are chosen such that each snapshot is represented by only one centroid. The centroid that represents each snapshot is the one that is the closest to the snapshot based on a distance measure. Here, the L$_2$-norm in the $\mathbb{R}^{N_p}$ space is used to compute this distance, which is given by 
        \[
        d_{i,k} = \sqrt{\sum_{j = 1}^{N_p} (S_i^j - C_k^j)^2},
        \tag{1}
        \label{eq:distance}
        \]
where $d_{i,k}$ represents the L$_2$-norm between the $i$-th snapshot and $k$-th centroid. The superscript $j$ in Eq.~\ref{eq:distance} indexes the number of dimensions $N_p$, or pixels, in the centroid/snapshot vectors.

\begin{figure}
    \centering
    \includegraphics[width = 0.5\columnwidth]{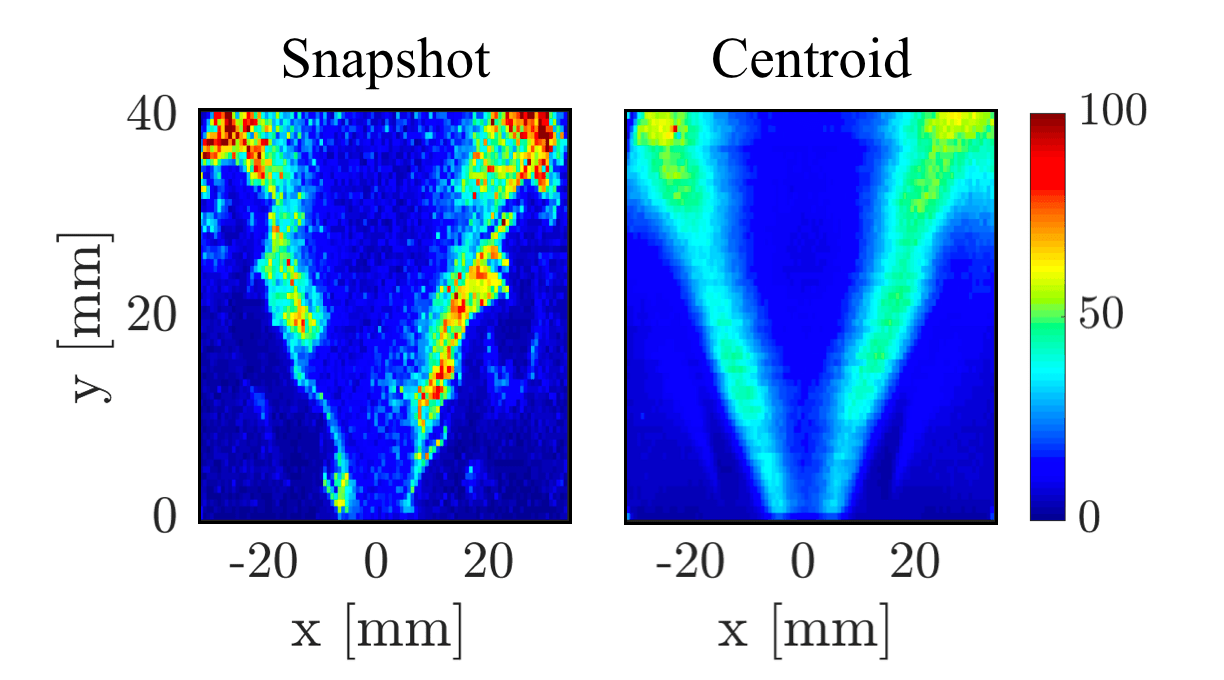}
    \vspace{-10pt}
    \caption{An example of a PLIF snapshot of an attached flame with its corresponding centroid.}
    \label{fig:centroid_snapshot_comparison}
 \end{figure}

This snapshot-centroid assignment is represented in an association matrix,
        \[
        T_{i,k} = \begin{cases} 1 & \text{if } S_i \in \mathbb{C}_k, \\
                                0 & \text{otherwise.} \end{cases}
                                \tag{2}
                                \label{eq:T_matrix}
        \]
Depending on the flow pattern represented by each centroid, the average distance of the snapshots assigned to each centroid can vary. Ideally, each centroid $C_i$ should be as close as possible to its assigned snapshots such that the centroid actually represents these snapshots. Note that the centroids are not flow fields that can be observed, but statistical patterns that approximate what can be observed. An illustration of this statistical effect is shown in Fig.~\ref{fig:centroid_snapshot_comparison}, which juxtaposes an example PLIF snapshot with its corresponding centroid. Here, the centroids are chosen using a k-means clustering algorithm \cite{arthur_kmeans}, a very common classification technique used in many unsupervised machine learning frameworks. 

In any k-means algorithm, centroid convergence rates are highly dependent on initial centroid locations as well as the input number of centroids, $N_k$. Here, the initial centroid locations are determined using the k-means++ routine, a variant of k-means as used in Ref.~\cite{kaiser_crom} and described in Ref.~\cite{arthur_kmeans}. Instead of picking the initial locations arbitrarily, which may lead to some centroids occupying very similar regions of the phase space initially, k-means++ initializes with the goal of achieving a large amount of separation between the centroids. The algorithm is summarized as follows. The first centroid, $C_1$, is randomly assigned to a snapshot $S_i$ in $\cal S$. Then, $C_2$ is assigned to another snapshot with probability proportional to the squared distance from its closest centroid. This initialization progresses for all centroids up to $C_{N_k}$, and ensures that the chance of choosing an initial centroid location far from an already existing centroid is high. Note that the k-means++ initialization is a stochastic algorithm. Therefore, multiple realizations of the algorithm are necessary to assess the predictive capabilities of the method in a statistically meaningful way.

With the centroid initialization method outlined, the full k-means algorithm as detailed in Ref.~\cite{arthur_kmeans} is summarized as follows:
    \begin{enumerate}
      \item Determine the initial distribution of $N_k$ centroids ${\cal C}$ with k-means++.
      \item Assign all $N$ snapshots to the nearest centroid as per Eq.~\ref{eq:distance}, accumulating the association matrix $T_{i,k}$ as per Eq.~\ref{eq:T_matrix}. 
      \item Update the k-th centroid by computing the center of mass of all the snapshots in the k-th cluster,
          \[
            C_k = \frac{\sum\limits_{i=1}^{N}S_i T_{i,k}}{\sum\limits_{i=1}^{N}T_{i,k}},\quad k = 1,\ ..., \ N_k, \quad i = 1, \ ...,\ N.
            \tag{3}
          \]
      \item Repeat (2) and (3) until convergence, where convergence is defined as the point at which an additional iteration would cause all centroids to change by a negligible distance. 
    \end{enumerate}
    
\subsection{Transition Matrix} 
\label{sec:crom:transition_matrix}
The prediction tool of the CROM approach is the transition matrix, which is the practical ROM necessary to make state forecasts. The transition matrix provides the probability of a snapshot transitioning from one cluster to another within a given forward time-step $\Delta t$. In a mathematical sense, this matrix represents a Markovian discrete time-step mapping. The transition probability of an image in cluster $j$ transitioning to an image in cluster $k$ is obtained based on the association matrix $T$ as:
    \[
    {\cal P}_{j,k} = \frac{\sum_{m=1}^{N-1} T_{m,j} T_{m+1,k}}{\sum_{m=1}^{N-1} T_{m,j}},  \ \text{for} \ k = 1,\ ..., \ N_k. 
    \tag{4}
    \label{eq:P_matrix}
    \]
The transition matrix ${\cal P}$ defined above is valid only for a finite time-step $\Delta t$. A key assumption regarding the cluster-based ROM framework is that of Markovianity, i.e., that the finite time-step transport of probability distributions using the transition matrix generated by the snapshots in set $\cal A$ is memoryless. Any finite-dimensional spatial discretization of Markovian governing equations will naturally yield a Markovian dynamical system. Here, it has only been assumed that the governing equations for the field measured are themselves memoryless. This is reasonable since fluid flow equations are memoryless. Moreover, given the large density of data points, it is reasonable to consider the system experimentally observed to follow Markovian dynamics as well. Model validation techniques for forecasting purposes are presented in Sec. \ref{sec:results-forecasting}. More details are provided in Ref.~\cite{kaiser_crom} (Sections 2.2.3 and 5).

Additionally, before centroid analysis and implementation of the transition matrix $\cal P$ in a prediction setting, it is useful to re-arrange the clusters in some probability-based order. One method (used in these results) is to order the clusters by descending eigenvalue modulus of $\cal P$ for clearer identification of cluster groups within the matrix structure \cite{kaiser_crom}.

\subsection{Forward Propagation Model} 
\label{sec:crom:forw_prop_model}
The centroid set ${\cal C}$ combined with the transition matrix ${\cal P}$ now serve as the prognostic model. From an initial PDF of the clusters $P_0 \in \mathbb{R}^{N_k}$ the forward model determines the probability distribution of centroids at a future time $t$:
    \[
    P_{t} = {\cal P}^{n_s} P_0,
    \tag{5}
    \label{eq:PDF_propagation}
    \]
where $n_s = t/\Delta t$. It is important to note that in the infinite time limit, the probability distribution represents the statistically stationary state that can be obtained from all the snapshots. This is the ergodic behavior of the transition matrix:
    \[
    \lim_{t \to \infty} P_t = {\bm e} \approx {\bm q}\text{, }~~ {q_k} = \frac{\sum_{m=1}^{N} T_{m,k}}{N}, \ \text{for} \ k = 1,\ ..., \ N_k.
    \tag{6}
    \label{eq:q_vector}
    \]
In Eq.~\ref{eq:q_vector}, $\bm e$ represents the PDF in the ergodic limit of the transition matrix, and $\bm q$ represents the initial cluster distribution of the snapshots. Furthermore, as $\cal P$ is a Markov matrix, its first eigenvector $\bm v_1$ corresponds to the distribution $\bm e$ (${\bm e} = {\bm v_1}$). Simply put, this feature states that the model gradually loses its predictive capability as $\cal P$ is raised to successively higher powers. Therefore, the transition matrix does not give information about any upcoming flame transition or flame state after a certain finite number of time-steps. This point can be defined as a finite critical time,  $\tau_h$, after which the probability distribution remains stationary regardless of the initial PDF $P_0$; this is referred to as the \textit{prediction horizon time} \cite{kaiser_crom}. As will be discussed in the results section, $\tau_h$ is highly sensitive to the type of data used to apply the CROM methodology, and is a key metric in evaluating the forecast power of one particular dataset over another. This will be the object of the discussion in Sec.~\ref{sec:results-forecasting}.

\subsection{Labeling of Centroids} 
\label{sec:labeling}
The clusters obtained from the CROM methodology isolate flow patterns in an unsupervised manner, but need to be assigned to states of interest for meaningful interpretation and forecasts. Here, it is explained how each cluster is labeled with the ``detached flame", ``attached flame" or ``transitioning flame" category. The centroid classification procedure is presented using a model generated with a combined dataset (OH-PLIF + PIV-x/y/z components). 

\begin{figure}[!tbp]
      \centering
      \subfloat[]{\includegraphics[width=0.48\columnwidth]{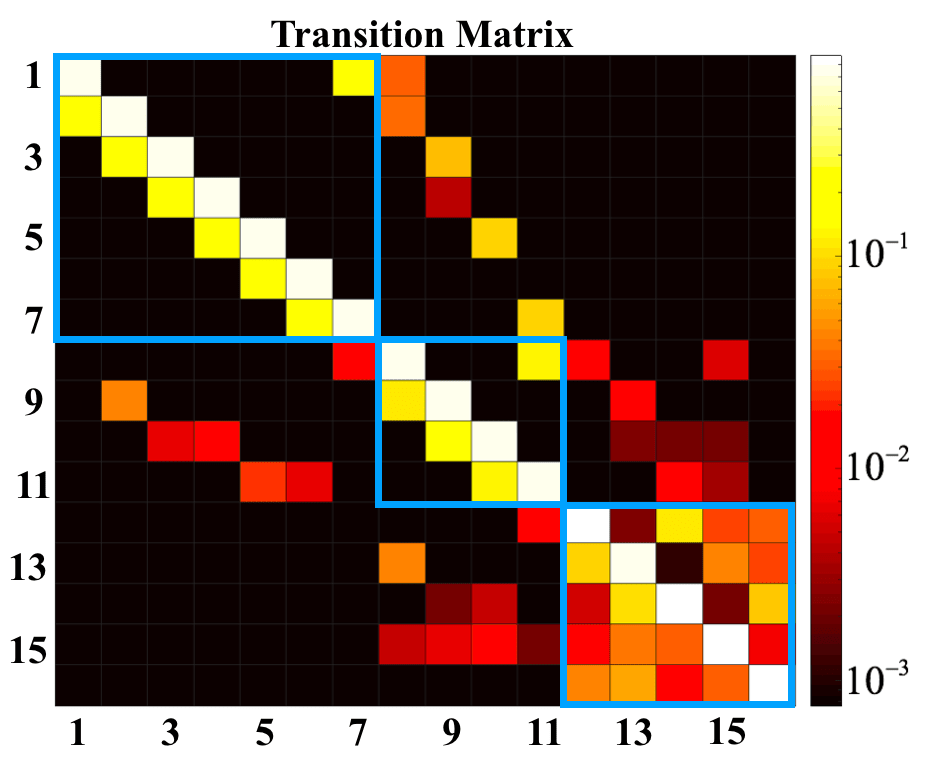}\label{fig:combined_transition_matrix}}
      \hfill
      \subfloat[]{\includegraphics[width=0.47\columnwidth]{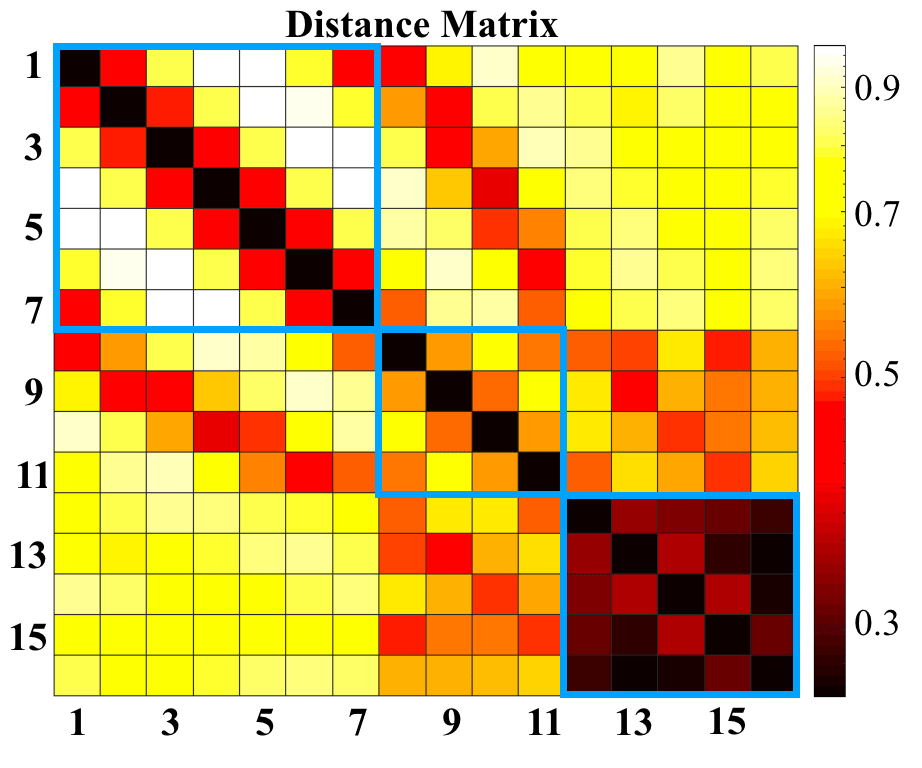}\label{fig:distance_matrix}}
      \caption{(a) Transition matrix with substructures boxed. (b) Distance matrix with substructures boxed.}
    \end{figure}

Figure~\ref{fig:combined_transition_matrix} displays the transition matrix with probabilities appearing as elements of the heatmap color-coded in log scale for $N_k = 16$. Three distinct structures are identified within the transition matrix as indicated by the blue boxes; further detail regarding the assignment of each structure to a physical state follows this list: 

    \begin{enumerate}
        \item Clusters 1-7 represent a periodic step-like probability structure, where the most likely path (aside from remaining in the same cluster) is to move on to the next cluster.
        \item Clusters 12-16 represent a highly interconnected probability structure (i.e. a snapshot in one cluster in this group has a roughly equally probable chance of moving on to any other cluster in this group, not just the next cluster).
        \item Clusters 8-11 are also step-like and periodic, similar to 1-7, but are probabilistically connected to both of the regions described in (1) and (2). 
    \end{enumerate}
    
Centroid groups are further identified via the \textit{cluster distance matrix} shown in Fig.~\ref{fig:distance_matrix}, which displays the L2 distance between each centroid \cite{kaiser_crom}. The distance matrix is symmetric with diagonal equal to zero. 

Centroids corresponding to the attached state are expected to be relatively similar in the phase space, as there should be far fewer possible phase space realizations of an attached flame in the given domain than a detached flame. This coincides with centroids 12-16 as indicated by the distance matrix, which is a grouping of centroids markedly close together. A clear cluster grouping is also visible when observing centroids 12-16 in the transition matrix -- the probabilistic interconnection of these centroids is related directly to their phase space similarity. Based on the above analysis, centroids 12-16 are associated with the attached flame state. Visualizing these centroids confirms the classification, as seen in Fig.\ref{fig:attached_combined}.

Of the two remaining groups (1-7 and 8-11), one must be assigned a ``detached" label and the other a ``transition" label. Centroids 8-11 in the transition matrix are connected to the other two groups: for example, a snapshot starting in clusters 8-11 has some probability of entering either clusters 1-7 or clusters 12-16. Centroids 8-11 are also identifiable as transition centroids via the distance matrix; centroids 1-7 are far from centroids 12-16, but centroids 8-11 are closer to both. Clusters 8-11 are therefore assigned the transition label. The corresponding centroids are plotted below in Fig.~\ref{fig:detached_combined}.

This leaves the detached label assignment for clusters 1-7. These clusters differ from the transition centroids since the matrix $\mathcal{P}$ does not allow transition to the attached centroids (12-16). Centroids 1-7 are plotted in Fig.~\ref{fig:detached_combined}, confirming the detached assignment. Note that there are more detached clusters than there are attached clusters, as is expected -- the detached flame state should occupy a larger region of the phase space than the attached state based on the experimental dataset.

\begin{figure}
    \centering
    \includegraphics[width = 0.8\columnwidth]{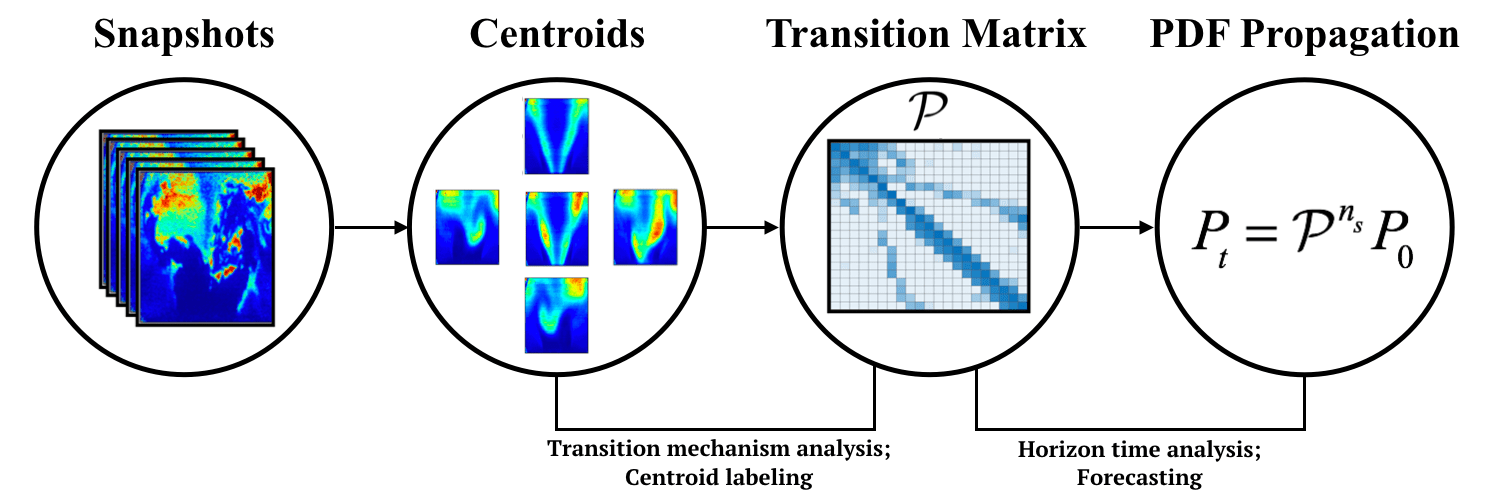}
    \caption{CROM workflow. The k-means clustering procedure occurs between the first two steps. Centroids and transition matrix shown here are arbitrary.}
    \label{fig:crom_model}
\end{figure}

CROM then provides the labeled centroids and the transition matrix, which can be used for a) analysis of the centroids and transition mechanism (Sec.~\ref{sec:results-centroids}), and b) prediction of flame transition and analysis of horizon time (Sec.~\ref{sec:results-forecasting}). In context of the current experimental setup, this approach is valid only when the system is ergodic. For additional perspective, a summary schematic of the model procedure is shown in Fig. \ref{fig:crom_model}.

\section{Analysis of the Transition Mechanism}
\label{sec:results-centroids}

Not only does the CROM methodology allow for the construction of a predictive model, but it also helps infer the physical mechanism through which particular events happen. In this case, CROM can be used to extract the sequence of events leading to flame detachment or attachment. The outcomes of CROM (centroids and transition matrix) are used in this section to perform this analysis. 

First, an ideal cluster number $N_k$ is chosen such that it is low enough to simplify the analysis but high enough to better identify all three flame states (Sec.~\ref{sec:num_cluster_transition}). Second, the clustering process is applied to the full dataset combining OH-PLIF and all the velocity components. As a result, each centroid obtained is composed of 4 different fields (OH and the three velocity directions). It will therefore be possible to describe the interaction between the flame and the velocity field during the flame transition process. Third, this transition process (both from detached to attached flame and attached to detached flame) is analyzed by combining the information provided by the transition matrix in the form of cluster transition probabilities with the centroids for the combined dataset. In particular, the most likely sequence of events leading to attachment or detachment will be extracted.

\subsection{Number of Clusters} 
\label{sec:num_cluster_transition}

\begin{figure}
    \centering
    \includegraphics[width = 0.8\columnwidth]{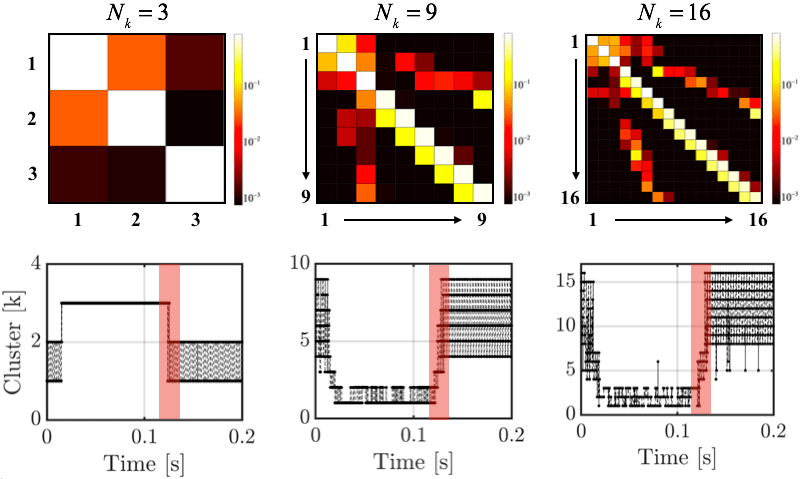}
    \vspace{-5pt}
    \caption{(Top) Transition matrices for $N_k=$ 3, 9, and 16. (Bottom) Corresponding cluster time series for the first 0.4 seconds.}
    \label{fig:matrix_comparisons}
 \end{figure}

For optimal mechanism analysis, it is critical to choose a cluster number that discretizes the phase space in such a way that all states of interest are captured. In this case, the goal is to choose a lower bound on $N_k$ such that the attached, detached, and transition flame states are properly refined. Transition matrix structures and cluster time series are compared in Fig.~\ref{fig:matrix_comparisons}. The \textit{cluster time series} plots (bottom row in Fig.~\ref{fig:matrix_comparisons}) show the time evolution of the images represented by associating each snapshot to its closest centroid. As seen from the data, the flame is initially in one state, but quickly transitions to another. Later in time (around $t=0.12$s), the combustor undergoes another transition and moves back to the original state. The red highlighted regions in Fig~\ref{fig:matrix_comparisons} show this second transition -- as the total number of clusters is increased, so too is the refinement of the transition mechanism. The number of clusters chosen is $N_k=16$, as it best captures the transition process and contains the necessary resolution of all three flame states of interest. The transition matrix structure shows clear groups that can be labeled as explained in Sec.~\ref{sec:labeling} -- clusters 1-3 are associated with an attached flame, clusters 8-16 with detached, and clusters 4-7 with a transition state. Figure~\ref{fig:matrix_comparisons} also implies that the selection of $N_k$ is more complex than assigning simply one cluster to one state (i.e. for three states, $N_k=3$), as the cluster number is dependent on the time spent in each state. Because the $N_k$ requirement depends on the sufficient resolution of the transition process, and the flame transition between attached/detached states occurs very quickly compared to the flame residence time in either attached or lifted states, in the end of the clustering process there should be more than one cluster per label. The transition matrix then has non-zero transition probability for several cluster changes, which allows variations in the density of trajectories in each individual flame state to be better represented. For example, having only one centroid in the detached state would be a drastic oversimplification of the evolution of trajectories corresponding to a detached flame, and would not capture the associated periodicity of the flame anchoring point. 

Besides facilitating the interpretation of the CROM output, using the smallest possible cluster number that captures the states of interest decreases the statistical uncertainty for the entries of the transition matrix. Therefore, it should be noted that if the sole interest is the forecasting power of the model, the optimal number of clusters $N_k$ will be different as will be shown in Sec.~\ref{sec:results-forecasting}. For the purposes of centroid analysis as it pertains to transition mechanism identification, in the following discussion, centroids were produced for a cluster discretization of $N_k=16$.
    
\subsection{Description of the Bistable State via Centroids}
\label{sec:centroid_analysis}

In this section, the centroids obtained for the value $N_k=16$ are analyzed to illuminate the behavior of the swirling flame during the attached, detached and transitioning phases.

\begin{figure}
      \centering
      \includegraphics[width = 0.8\columnwidth]{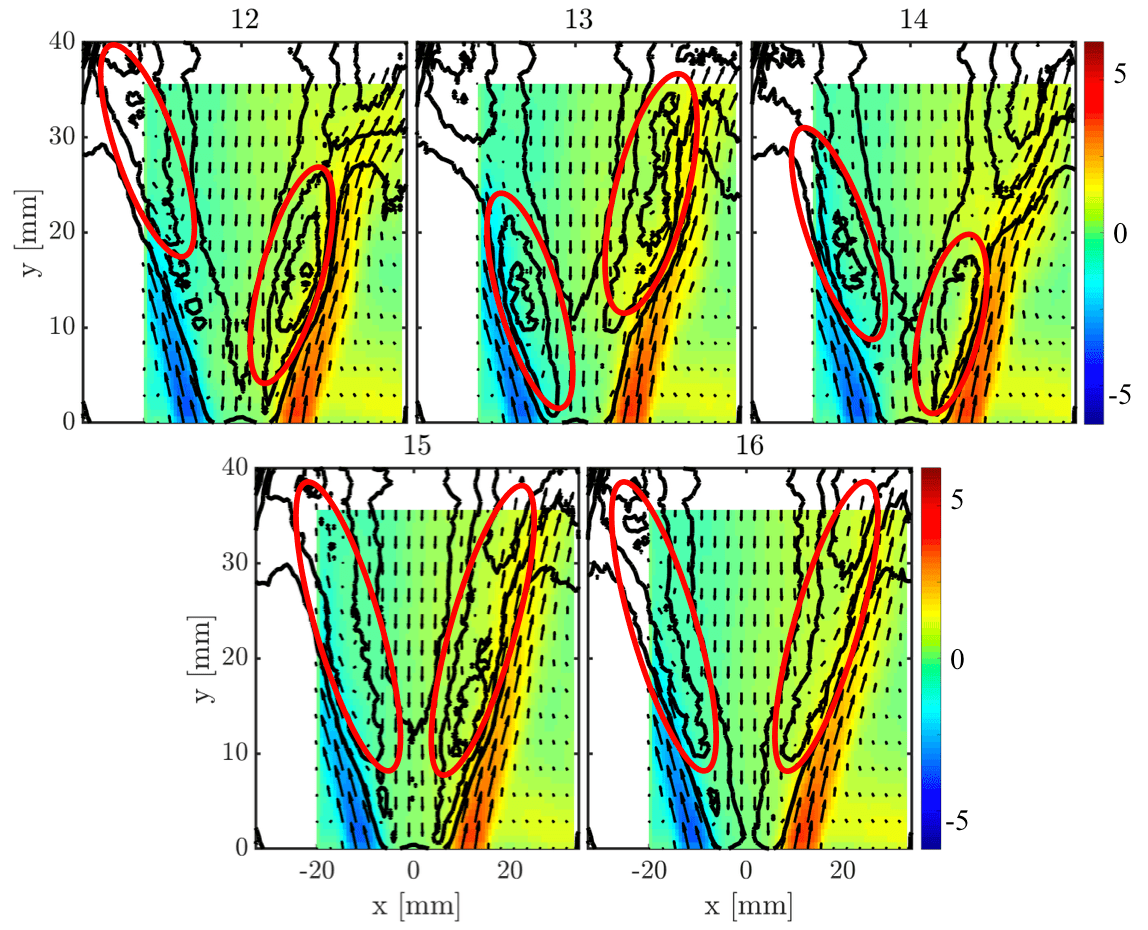}
      \caption{Attached centroids of the combined dataset. OH-PLIF isocontours indicated in black lines. PIV-x and y components given by arrow overlays, and PIV-z is given by the heatmap (colorbar units in m/s). Circled regions enclose regions of increased OH-concentration.}
      \label{fig:attached_combined}
\end{figure}

\begin{figure}
      \centering
      \includegraphics[width = 0.6\columnwidth]{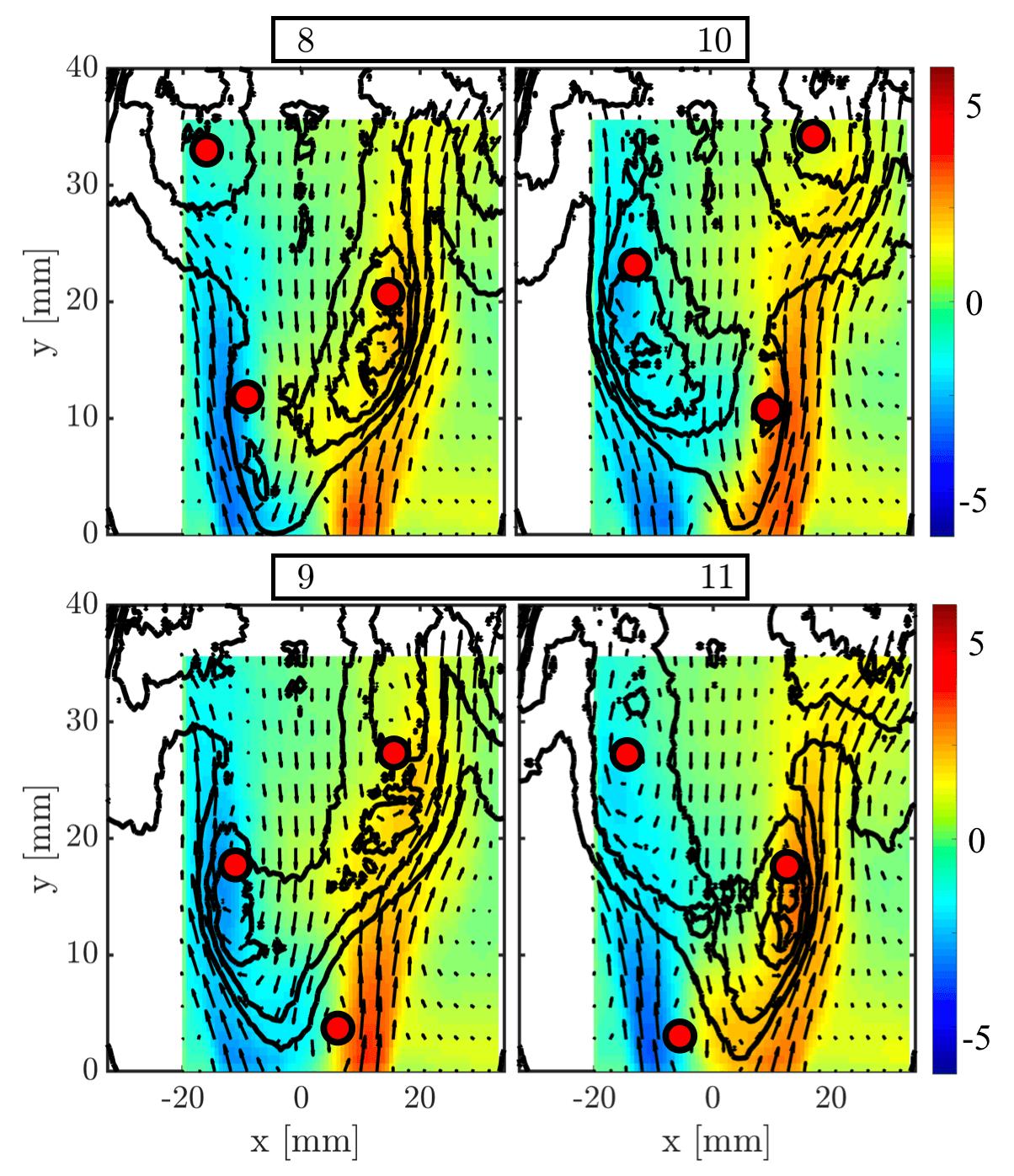}
      \caption{Transition centroids of the combined dataset. OH-PLIF isocontours indicated in black lines. PIV-x and y components given by arrow overlays, and PIV-z is given by the heatmap (colorbar units in m/s). Recirculation zone centers in indicated by markers.}
      \label{fig:transition_combined}
\end{figure}

\begin{figure}
      \centering
      \includegraphics[width = 1\columnwidth]{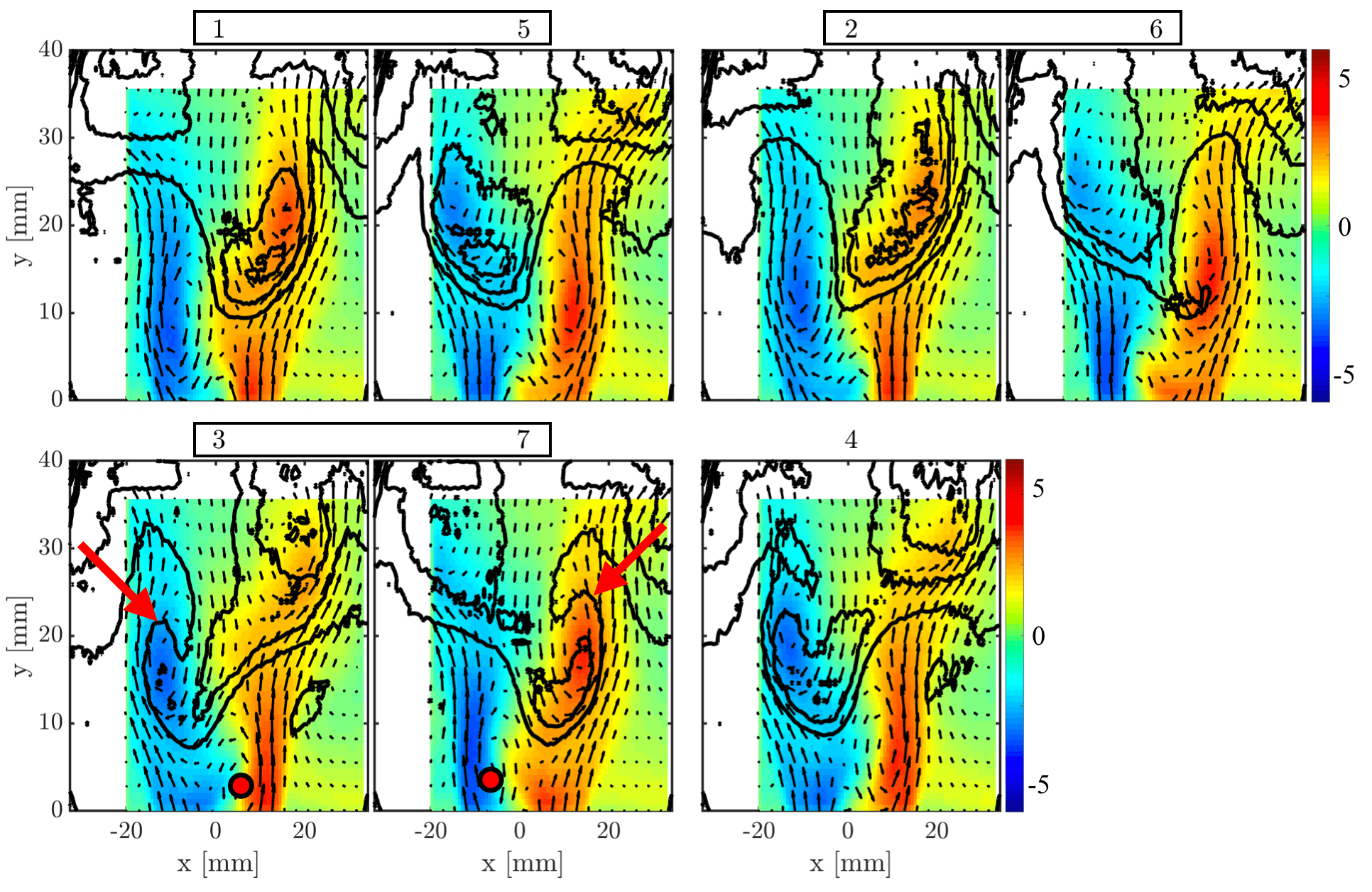}
      \caption{Detached centroids of the combined dataset. OH-PLIF isocontours indicated in black lines. PIV-x and y components given by arrow overlays, and PIV-z is given by the heatmap (colorbar units in m/s). In centroids 3 and 7, the hook-like structures are indicated by red arrows and the recirculation zones near the burner exit are marked in red.}
      \label{fig:detached_combined}
\end{figure}

The complete set of attached, transition, and detached centroids are shown in Figs. \ref{fig:attached_combined}, \ref{fig:transition_combined}, and \ref{fig:detached_combined} respectively for the full dataset. These images essentially show how the clustering algorithm organizes the raw snapshots. In the figures, OH isocontours are shown as black lines, and the arrows represent the velocity in the axial and transverse directions (x and y). Out-of-plane velocity (z component) is shown by the colored contour.

In the attached state, full flame symmetry is observed for centroids 15 and 16 which is expected -- the OH isocontours indicate a similar flame shape to the instantaneous PLIF attached image seen in Fig.~\ref{fig:schematic_and_flame}. However, in some attached centroids (12, 13, 14), there exists asymmetric alternating high-OH concentration regions (circled in red Fig.~\ref{fig:attached_combined}). For the same centroids, some slight velocity field asymmetries are also observed ("bending" of the out-of-plane velocity streaks and alternating recirculation zone presence), potentially indicating the initial formation of a PVC. This is also observed in experimental results \cite{oberleithner_pvc, qiang}.

Images corresponding to the transition centroids (8-11, Fig. \ref{fig:transition_combined}) show more complex structures for the velocity and OH profiles. These centroids show strong flame and flow asymmetry, suggesting that the PVC gained in strength. All transition centroids clearly display alternating recirculation zones (zig-zag structure as observed in \cite{oberleithner_pvc}) with coinciding high OH-concentration fields. It appears that the vortices creating vorticity normal to the measurement plane serve as flame anchoring points. The transition centroids obtained can be associated in pairs 8/10 and 9/11 which exhibit symmetry with respect to the swirling axis. The radial symmetry implies that the detachment process happens independently of the swirling process, but is localized in the azimuthal direction.

The detached centroids are shown in Fig. \ref{fig:detached_combined}. The OH concentration and the flow field are similar to those of the transition centroids, and are more pronounced. The flame front and the recirculation zones are in particular more lifted. One key difference that can be noted in terms of the flame topology is the existence of hook-like structures for the OH contour in centroid 3 and 7 (red arrows). This suggests that the flame is exposed to stronger strain rates than during the transition. 

A key aspect from the above discussion is that a pure phase-space clustering process is able to isolate important flow features relevant to the transition phenomenon in an unsupervised manner.

\subsection{Analysis of the Bistable Transition via Transition Matrix}
\label{sec:transition_matrix_analysis}

In this section, two different transition mechanisms are analyzed: the detachment process (attached to detached flame) and the attachment process (detached to attached flame). This section combines the physical information gained from individual centroid analysis (Sec~\ref{sec:centroid_analysis}) with the probability transition information from the transition matrix in Fig.~\ref{fig:combined_transition_matrix} in order to interpret the dynamics of the transition mechanism. The analysis below is conducted for model generated with the full dataset (OH and all three velocity field components). 

\subsubsection{Flame Detachment Process}

\begin{figure}
      \centering
      \includegraphics[width = 1\columnwidth]{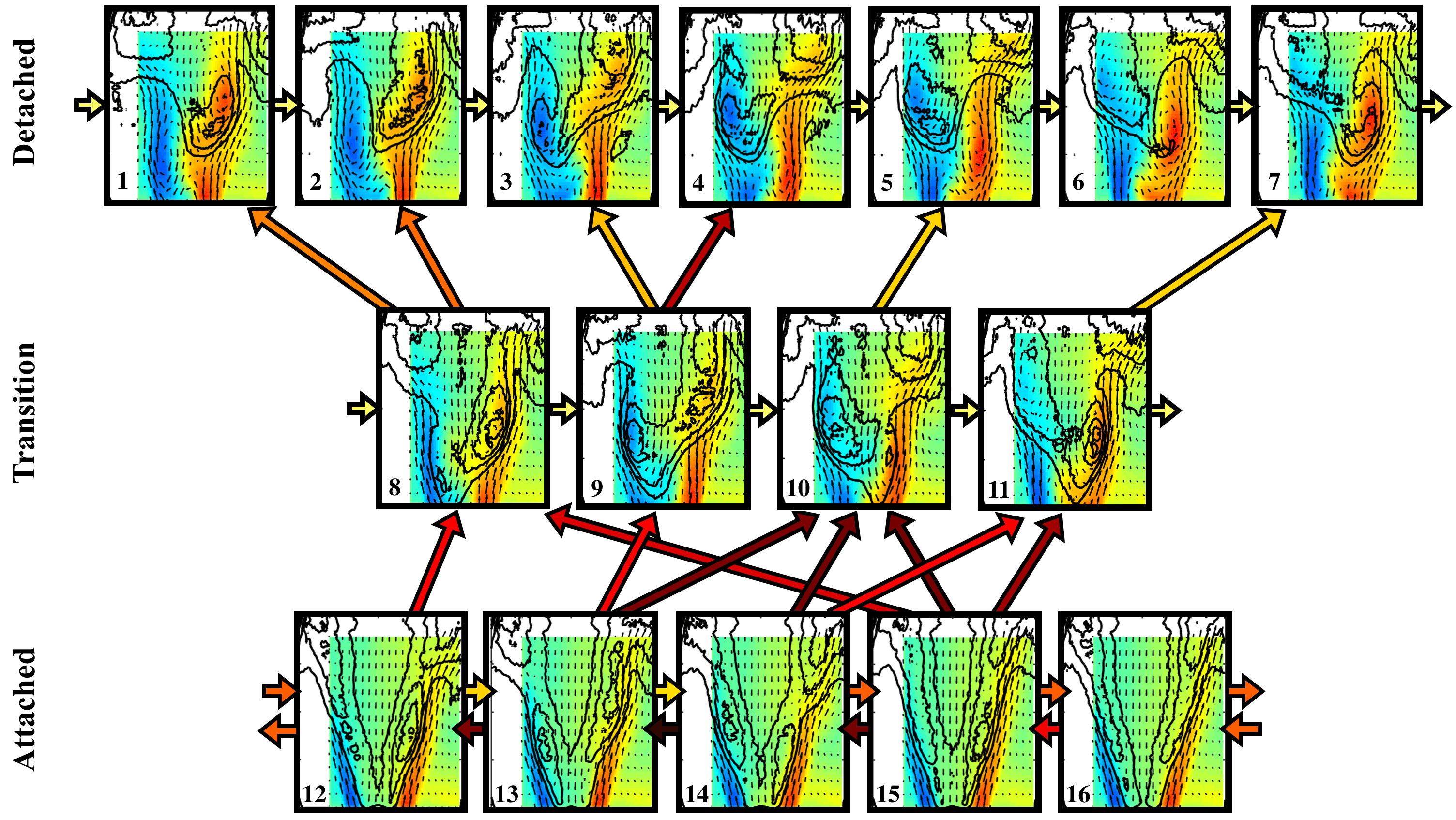}
      \caption{Probability paths for the cluster transitions in the flame detachment process. Arrows indicate the paths and are color-coded with the same colorbar as the transition matrix; darker colors are smaller probabilities, and brighter colors are higher probabilities.}
      \label{fig:detachment_process}
    \end{figure}

Figure \ref{fig:detachment_process} displays the probability paths as outputted by the model for detachment. Colors associated with the arrows are coded in the same scale as the probabilities in Fig.~\ref{fig:combined_transition_matrix} -- lighter colored arrows indicate higher probabilities. In Fig.~\ref{fig:detachment_process}, an interesting distinction is that the attached centroids associated with slight asymmetry in the OH field (centroids 12, 13, 14 and 15) have a higher chance of moving on to the transition clusters. Interestingly, the attached centroid that is most symmetric (centroid 16) has zero chance of moving on directly to a transition centroid, implying that asymmetry is a leading indicator of detachment. The velocity field asymmetry was also used as a marker of the beginning of flame detachment in the study of Oberleithner et al.~\cite{oberleithner_pvc}. Here, the results suggest that slight asymmetry of the flame itself could be at the inception of the detachment. Further analysis will be required to clearly identify the causes of the detachment. Nevertheless, this finding highlights the capabilities of CROM in assisting the interpretation of experimental data for transient flows.

The transition centroids are characterized by the tendency to evolve into a neighboring centroid. Each cluster appears to resolve a particular phase of the periodic swirling motion. Centroid 8 tends to evolve into 9, 9 into 10, 10 into 11, and 11 back into 8, etc. In the case of a detachment, both centroids 10 and 11 are likely to evolve into the detached centroids 5 and 7. Centroids 8 and 9 can each evolve into different detached centroids: 1/2 and 3/4 respectively. Similar features are seen within transition centroid pairs in the evolution to the lifted state; for example, centroids 9 and 11 (one mirrored pair) each evolve into detached centroids which exhibit a hook-like feature on the same side of the symmetry plane. Furthermore, centroids 8 and 10 (the other transition centroid mirrored pair) evolve into lifted centroids which depict the flame on the opposing side of the symmetry plane (i.e. centroid 8 is left-leaning and evolves into a right-leaning lifted flame -- vice-versa for centroid 10). The detached centroids depict a similar periodic tendency as the transition centroids, as the most probable path is to simply evolve into the next centroid. Assessment of the lower stagnation point in the detached regime gives insight to the PVC structure location, and it can be seen that this feature oscillates back and forth across the symmetry plane. An interesting feature to note is that the detachment process does not include centroid 6, which may be a product of the chosen cluster number.

\subsubsection{Flame Attachment Process}

\begin{figure}
      \centering
      \includegraphics[width = 1\columnwidth]{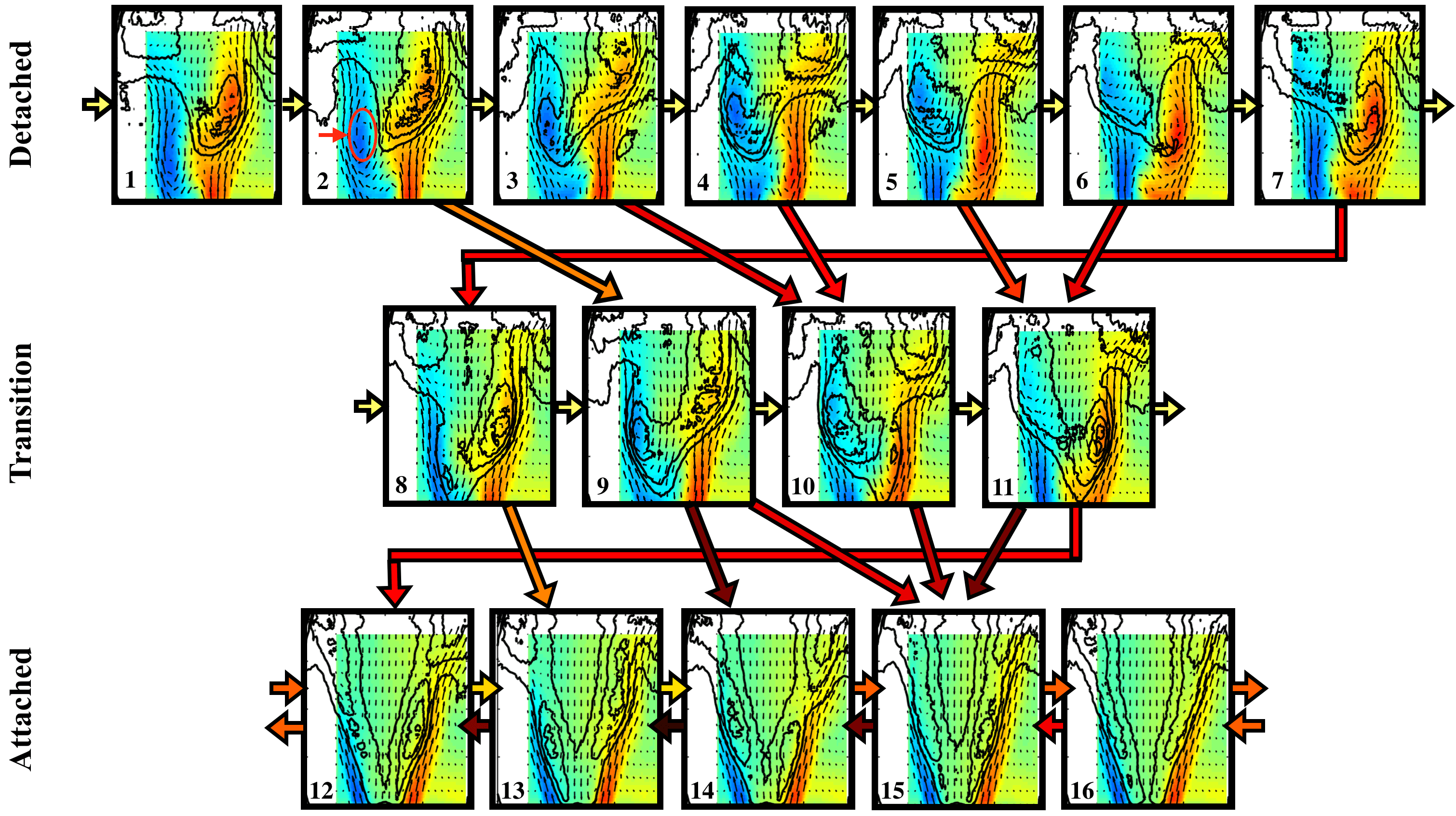}
      \caption{Probability paths for the cluster transitions in the flame attachment process. Arrows indicate the paths and are color-coded with the same colorbar as the transition matrix; darker colors are smaller probabilities, and brighter colors are higher probabilities.}
      \label{fig:attachment_process}
    \end{figure}

In the flame attachment case as outlined in Fig.~\ref{fig:attachment_process}, it is seen that a snapshot in the first cluster has no chance to directly enter the transition region. The most likely detached-to-transition pathway is given by the progression from centroid 2 (right-leaning detached flame) to centroid 9. Centroid 2 differs from the other detached centroids in that there exists a strong recirculation zone on the left of the image (highlighted in red) almost at the same axial location as the flame anchoring point on the right. The flame can therefore be entrained toward the nozzle by the local negative axial velocity. 

It is noted that centroids 1 and 2 exhibit a very similar flame front but have a very different velocity field. Centroid 1 does not play a direct role in the attachment mechanism while centroid 2 is key for the inception of the attachment. The velocity field is therefore not only at the root of the flame detachment, but also plays a key role in flame attachment. In the case that centroid 2 evolves into centroid 9, development of a recirculation zone very near the nozzle exit is seen. In fact, higher probability cases in which a flame enters the transition state (e.g. centroids 4 to 10,  5 to 11) all depict the presence of a lower recirculation zone near the nozzle. The most probable path for transition-to-attached is the progression from centroid 8 to centroid 13. Centroid 8 is the most symmetric centroid of all the transition clusters in terms of both the velocity and OH concentration. This appears to help the flame stabilize into the "V-shape" profile characteristic of the attached configuration.

\section{Prediction of Flame Transition}
\label{sec:results-forecasting}

Along with providing information about the transition mechanism, the transition matrix can also be used as a forecasting tool. The quantification of model predictability (or predictive strength) is found in the prediction horizon time, $\tau_h$, defined in Sec.~\ref{sec:crom}. It is the time after which the transition matrix converges to the ergodic probability distribution $\bm e$. This convergence is not a sudden process, since the transition matrix continuously diffuses towards its stationary state as it is raised to successively higher powers. Therefore, any input probability distribution will eventually converge to $\bm e$ given a large enough number of time-steps. 

Similar to Sec.~\ref{sec:results-centroids}, it is necessary to choose a value for parameter $N_k$, the number of clusters. Here, $N_k$ is chosen to optimize \textit{forecasting or predictive purposes}. With different constraints than in Sec.~\ref{sec:results-centroids}, it will be seen that the optimal number of cluster will also be different. It should be noted that in this section, the primary focus is to compare the predictive power of different types of data, where the data types in this case are OH-PLIF, PIV-x, y, and z. This means that unlike in Sec.~\ref{sec:results-centroids}, which utilized a single model composed of a combined dataset to assess the physical mechanisms of the transition, different models will be created for each type of data to compare horizon times, uncertainty and predictive capability.

This section is organized in the following manner. First, the process of extracting the horizon time $\tau_h$ from the transition matrix is demonstrated in Sec.~\ref{sec:horizon_time_determination}. Then, the procedure used to set the optimal number of clusters is described in Sec.~\ref{sec:number_of_clusters_prediction}. In Sec.~\ref{sec:horizon_time_comparisons}, horizon times obtained with different datasets are compared. Finally, in Sec.~\ref{sec:predictions}, flame transition forecasts are compared across models constructed with different data-types (PIV measurements and OH-PLIF). Model validation tests are shown in the Appendix.

\subsection{Determination of Prediction Horizon Time}
\label{sec:horizon_time_determination}

\begin{figure}
      \centering
      \includegraphics[scale=0.25]{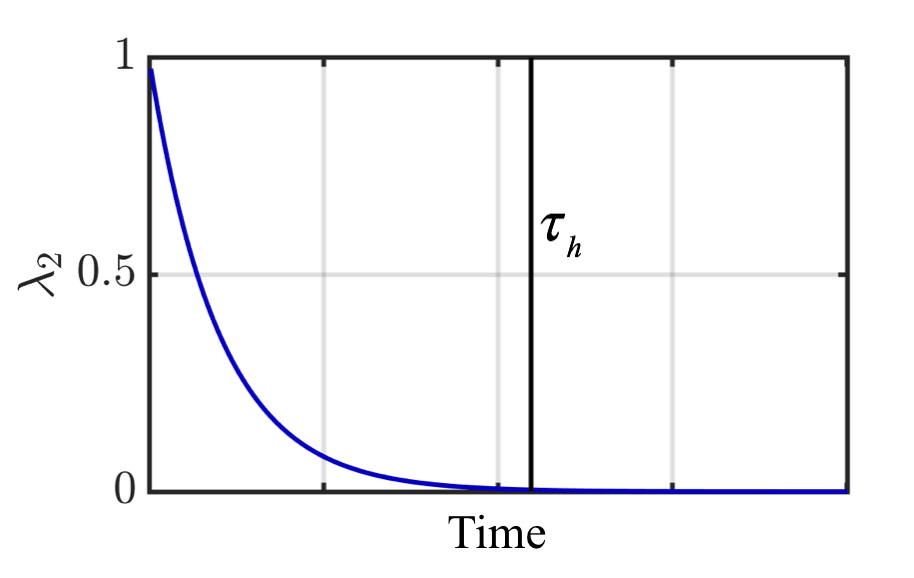}
      \vspace{-10pt}
      \caption{Example of second-largest eigenvalue convergence for some $\cal P$. The horizon time, $\tau_h$, is the time at which the slope of $\lambda_2$ falls below some threshold $\varepsilon$.}
      \label{fig:prediction_horizon_example}
\end{figure}

The prediction capability of the transition matrix can be quantified by the prediction horizon time $\tau_h$ (see Sec.~\ref{sec:crom}). There are several methods outlined in Ref.~\cite{kaiser_crom} that can be used to obtain $\tau_h$ for a given transition matrix $\cal P$ (second eigenvalue convergence to zero, probability distribution convergence to ergodic distribution, convergence of finite-time Lyapunov exponent, and Kullback-Leibler entropy convergence); here, the eigenvalue spectrum of $\mathcal{P}^n$ is monitored as $n$ increases. An inherent property of $\cal P$ is that the largest eigenvalue is unity and is stationary (does not change in time), meaning that an eigendecomposition of the transition matrix raised to any discrete positive power always yields a maximum eigenvalue of one. The corresponding eigenvector of the unity eigenvalue is the statistically stationary state (the ergodic limit). As the system evolves in time, the remaining $N_k-1$ eigenvalues gradually converge zero (this is a property of Markov chains). The second largest eigenvalue $\lambda_2$ is the last eigenvalue to reach zero. Its value is tracked as the transition matrix is raised to higher and higher powers. Figure~\ref{fig:prediction_horizon_example} shows a typical evolution of the second-largest eigenvalue with respect to time. In practice, the prediction horizon time is defined as the time at which $d \lambda_2/dt < \varepsilon$. Here, $\varepsilon=1\text{e}{-5}$.

\subsection{Number of Clusters}
\label{sec:number_of_clusters_prediction}

Similar to what was done for the analysis of the transition mechanism in Sec.~\ref{sec:results-centroids}, the cluster number must be carefully chosen. In this section, the purpose of CROM is different than in Sec.~\ref{sec:results-centroids}, and as such the optimal number of clusters/centroids $N_k$ is determined differently. Here, the goal is to optimize the prediction horizon time $\tau_h$ while decreasing the statistical uncertainty of the transition matrix entries. The measure of this statistical uncertainty is explained below.

Consider the square transition matrix $\cal P$, where each entry of the matrix is $\mathcal{P}_{i,j}$ and $i,j = 1,\ ..., \ N_k$. One can then associate a relative error to each element $\mathcal{P}_{i,j}$ as
 \[
    \delta \mathcal{P}_{i,j} = \sqrt{\frac{1 - \mathcal{P}_{i,j}}{N \mathcal{P}_{i,j}}},
    \tag{7}
 \]
where N is the total number of snapshots in the dataset. The measure chosen for the uncertainty quantity is the Frobenius norm of the relative uncertainty matrix, 
 \[
    ||\delta \mathcal{P}|| = \sqrt{\frac{1}{N} \sum_{i=1}^{N_k} \sum_{j=1}^{N_k} \frac{1 - \mathcal{P}_{i,j}}{\mathcal{P}_{i,j}}}.
    \tag{8}
 \]

In the rest of the section, the transition matrices and the clusters are constructed with data in the range [0s, 1s] of the available measurement. This training set covers 4 of the 8 flame transitions available in the data. The last two transitions (range [1s, 1.5s]) are kept to test the model performances (testing set). Transition matrices are constructed for different number of clusters in order to find the optimal $N_k$ sought here. 

\begin{figure}
      \centering
      \includegraphics[width=0.8\columnwidth]{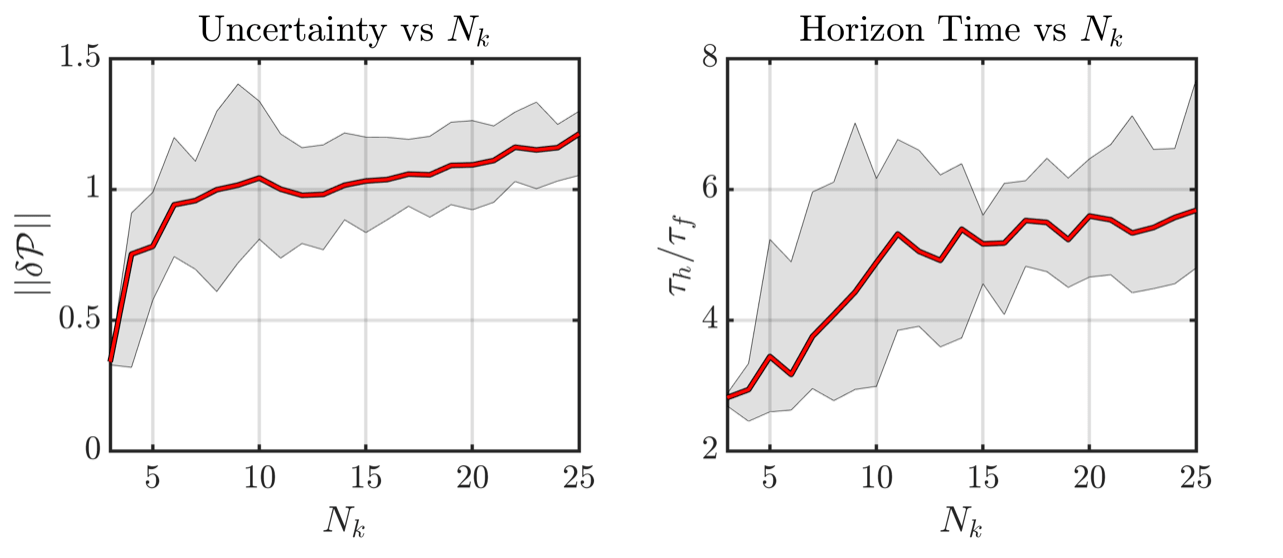}
      \vspace{-10pt}
      \caption{Out-of-plane velocity field transition matrix uncertainties (left) and horizon times (right) versus $N_k$. Maximum and minimum bounds are indicated by shaded boundaries derived from individual k-means++ runs, red lines indicate mean.}
      \label{fig:optimal_cluster_pivz}
\end{figure}

Figure \ref{fig:optimal_cluster_pivz} displays the uncertainty measure and normalized prediction horizon time as a function of cluster number, $N_k$, for the PIV-z (out-of-plane velocity) dataset. Prediction horizon time $\tau_h$ was normalized with the combustor flow-through time $\tau_f = 14.6$~ms. Because the k-means++ algorithm is stochastic (see Sec. \ref{sec:crom}), 20 realizations were run to ensure that the conclusions are meaningful. The enclosed shaded regions in Fig.~\ref{fig:optimal_cluster_pivz} represent maximum and minimum boundaries obtained from individual runs of the clustering algorithm, and the red line indicates the average of all runs.

The uncertainty metric growth rate in Fig.~\ref{fig:optimal_cluster_pivz} decreases dramatically after an initial increase until $N_k=6$. The error metric appears to be increasing, albeit slowly, beyond a small relative maximum at $N_k=10$ after the sharp increase at lower cluster numbers. A similar sharp increase trend is seen in the normalized prediction horizon plot, where the mean stabilizes after $N_k=10 \sim 11$. The results of both the prediction horizon and the matrix uncertainty suggest that the optimal cluster number for the PIV-z dataset could be $N_k =  11$. 

\subsection{Horizon Time Comparisons}
\label{sec:horizon_time_comparisons}
The process of choosing an optimal cluster number based on horizon time and transition matrix uncertainty was conducted for the other data-types in the same manner as shown for the PIV-z dataset in Sec. \ref{sec:number_of_clusters_prediction}. In the end of the process, different optimal cluster numbers $N_k$ are found for different data types. The results are summarized in Table~\ref{table:data_comparison}. Similar model validation test trends as shown for PIV-z in the Appendix were observed for models derived from all data-types. 

\begin{table}
      \centering
      \includegraphics[width=0.8\columnwidth]{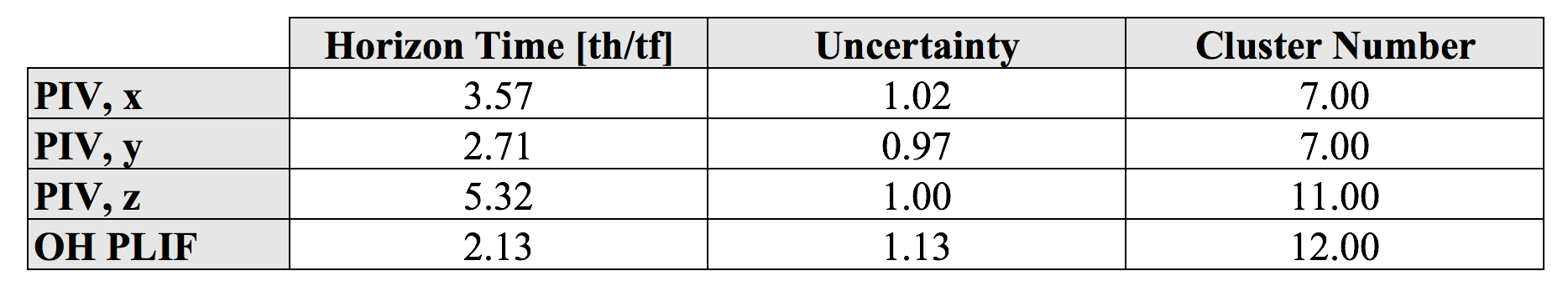}
      \caption{Optimal $N_k$ values with corresponding uncertainties and horizon times as extracted from Fig.\ref{fig:horizon_time_comparisons}.}
      \label{table:data_comparison}
\end{table}

\begin{figure}
      \centering
      \includegraphics[width=0.9\columnwidth]{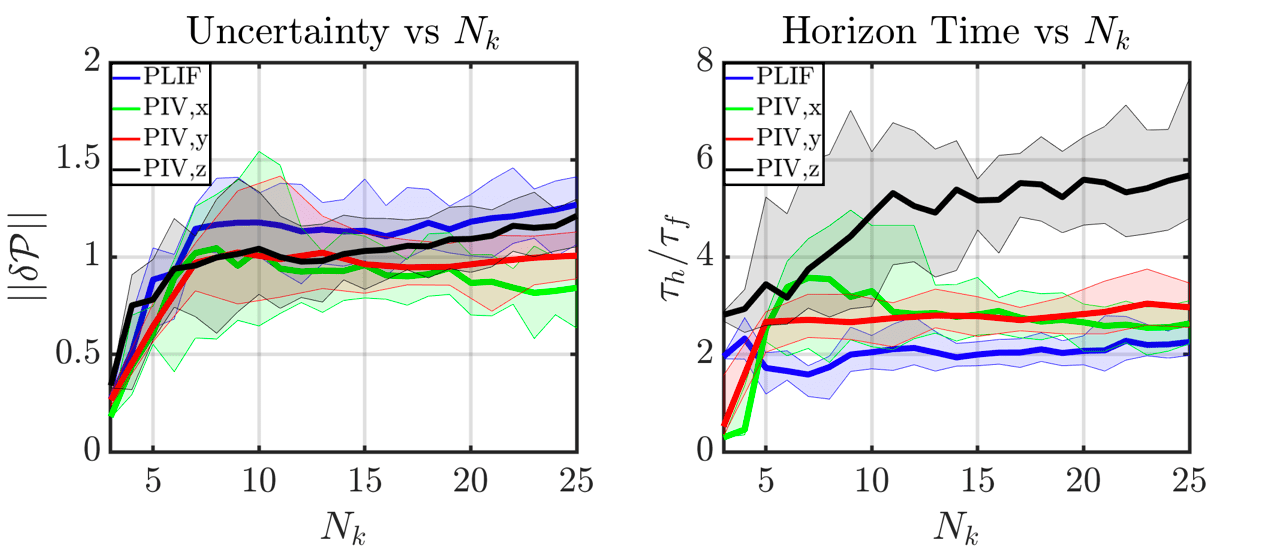}
      \vspace{-10pt}
      \caption{Comparisons of transition matrix uncertainty (left) and horizon times (right) for the four tested data types. The curves are averaged quantities from 20 independent k-means++ runs on the dataset. Maximum and minimum bounds are indicated by shaded boundaries.}
      \label{fig:horizon_time_comparisons}
\end{figure}

Figure~\ref{fig:horizon_time_comparisons} shows a more detailed comparisons of normalized horizon times and uncertainty measures for the four different datasets (OH-PLIF, PIV-x, y, and z directions) as a function of cluster number -- this is the same plot as presented in Fig.~\ref{fig:optimal_cluster_pivz}, but with overlays for all data types for clearer trend visualization. The curves shown are average quantities of 20 runs for each data type with minimum and maximum bounds for each curve displayed in the same corresponding color.

The plot shows  that the out-of-plane component of velocity (PIV-z) provides the highest horizon time for nearly all cluster numbers. In contrast, the OH-PLIF dataset interestingly gives the lowest horizon time for most of the tested cluster numbers. This is counter-intuitive since the OH concentration is a marker of the flame position and can be expected to be a good descriptor of the lift-off or attached states. This finding can be interpreted by considering the cause and effect of flame transition. The OH signal could be an effect of the flame transitioning to a new state, while the out-of-plane velocity field could be a cause. As a results, the PIV-z data contains more detail about the future state of the combustor. This agrees with prior analyses in that the PIV-z data was actually found to be at the root of flame transitions \cite{qiang,qiangaiaa}. The flame lift-off, for example, is aided by high-strain rates in the inner recirculation zone, which causes flame extinction and formation of a PVC. Since a signature of the PVC can be found in the out-of-plane velocity component (PIV-z), it is unsurprising that the predictive power of this field is high compared to other measurements. 

\subsection{Forward State Predictions}
\label{sec:predictions}

In practice, a tool to forecast transitions between different macroscopic states is invaluable to anticipate and control flame instabilities. As discussed in Sec.~\ref{sec:crom}, CROM can be used to predict the future probability distribution of the combustor state in terms of the clusters. Using the cluster labeling method described in Sec.~\ref{sec:labeling}, the predictions can be formulated in terms of the macroscopic flame states. Starting from a state where the flame is in either the detached or attached state, the goal is to understand if the model can accurately predict the probability of exiting the initial state, or transitioning. In other words, the objective is to determine how accurately the model can predict a flame detachment (flame exiting the attached state) as well as a flame attachment (flame exiting the detached state). The data was partitioned in training and testing sets in the same way as indicated in Sec.~\ref{sec:number_of_clusters_prediction}. The state prediction process is conducted as follows: 

\begin{enumerate}[label=(\alph*)]
    \item Pick one cluster labeled as attached (resp. detached). Start with a $\delta-$distribution $P_0$ in this particular cluster. $P_0$ is a vector equal to $0$ everywhere and $1$ for the index of the cluster picked. This situation would correspond to observing the flame as being in a state that corresponds to the cluster picked.
    \item Determine all snapshots associated with the chosen attached (resp. detached) centroid in (a) in the cluster time series for the testing dataset.
    \item Advance $P_0$ to $P_{\tau_p}$ using the transition matrix (Eq.~\ref{eq:PDF_propagation}), where $\tau_p$ is the time in the future at which predictions are sought.
    \item Use the experimental snapshots at time $\tau_p$ relative to the original snapshots determined in (b) to determine an experimentally obtained $P_{\tau_p,exp}$ from the testing dataset.
\end{enumerate}

\begin{figure}
      \centering
      \includegraphics[width=1\columnwidth]{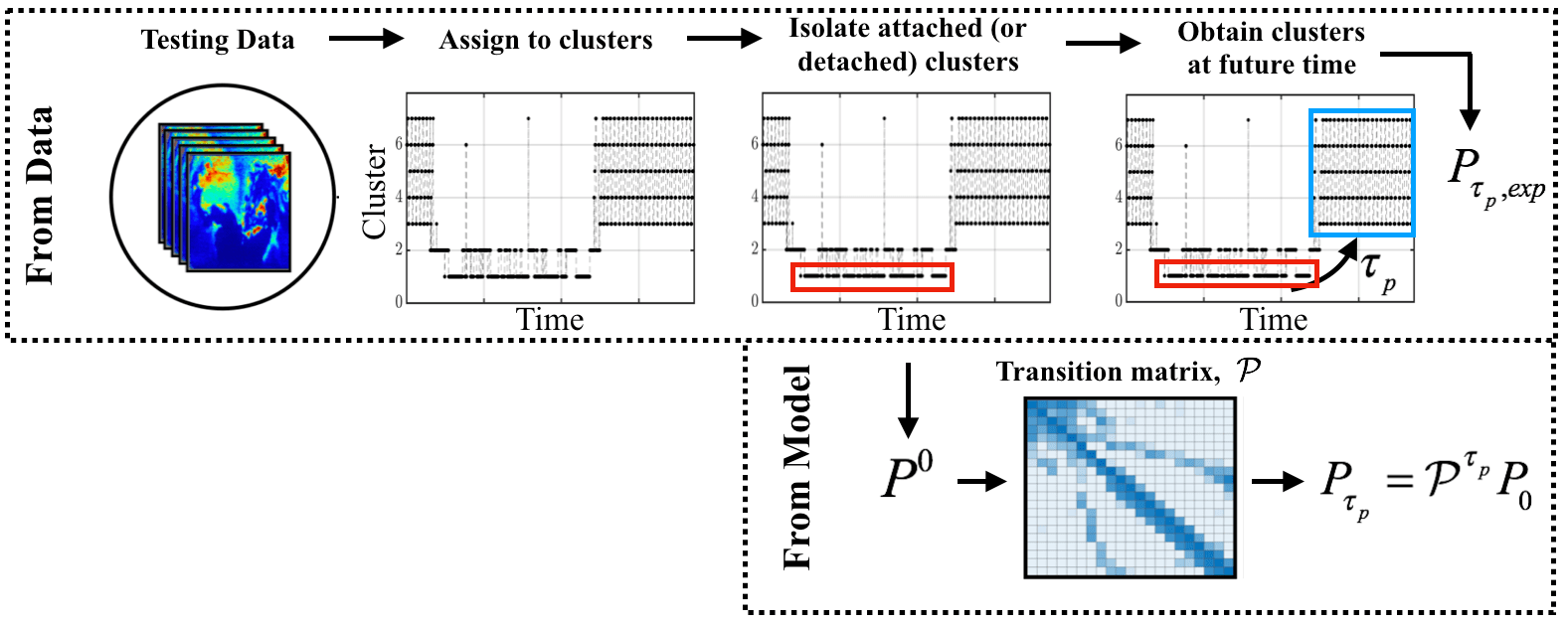}
      \caption{Schematic juxtaposing the procedure for finding $P_{\tau_p,exp}$ (directly from data) and the procedure for for finding $P_{\tau_p}$ (from CROM).}
      \label{fig:exact_pdf_method}
\end{figure}

The computed (from model, $P_{\tau_p,exp}$) and experimental (from data, $P_{\tau_p}$) probabilities are then compared to determine the accuracy of the ROM. This is schematically shown in Fig.~\ref{fig:exact_pdf_method}.

Forecasts are compared for out-of-plane velocity field (PIV-z) and OH-PLIF measurements for conciseness. The number of cluster used for each data-type is indicated in Table~\ref{table:data_comparison}. Two prediction scenarios were tested for each data-type: 1) the snapshot initially resides in the attached state (100\% chance of a snapshot in attached cluster), and 2) the snapshot initially resides in the detached state (100\% chance of a snapshot in detached cluster). For each scenario, three prediction times were assigned to illustrate model performance at various time scales: $\tau_p = \tau_f = 14.6$~ms (one flow-through time),  $\tau_p = 0.1\tau_f = 1.46$~ms (short-time prediction), and $\tau_p = 7.0\tau_f = 10.22$~ms (long-time prediction, i.e. beyond the prediction horizons for both PIV-z and OH-PLIF). The initial cluster distributions $P_0$ were propagated through the transition matrix for these specified $\tau_p$ values, and the resulting final distribution was then compared with that of the testing dataset.

Because different models are generated when using PIV-z and PLIF data (different cluster numbers, different clusters, different transition matrix), the comparison of performance between datasets is non-trivial. Here, the initial cluster for each dataset is chosen such that 1) the cluster appears to depict a similar flow state, and 2) the final probability distributions $P_{\tau_p,exp}$ are similar for both model outputs. This ensures that both clusters are representative of a similar combustor state. Therefore, the test compares the ability of both datasets to predict transitions that are as close as possible. This criterion becomes more apparent in the results below. This method was used with different pairs of clusters showing similar results. Here it is shown for one of these pairs.

\begin{figure}
      \centering
      \includegraphics[width=1\columnwidth]{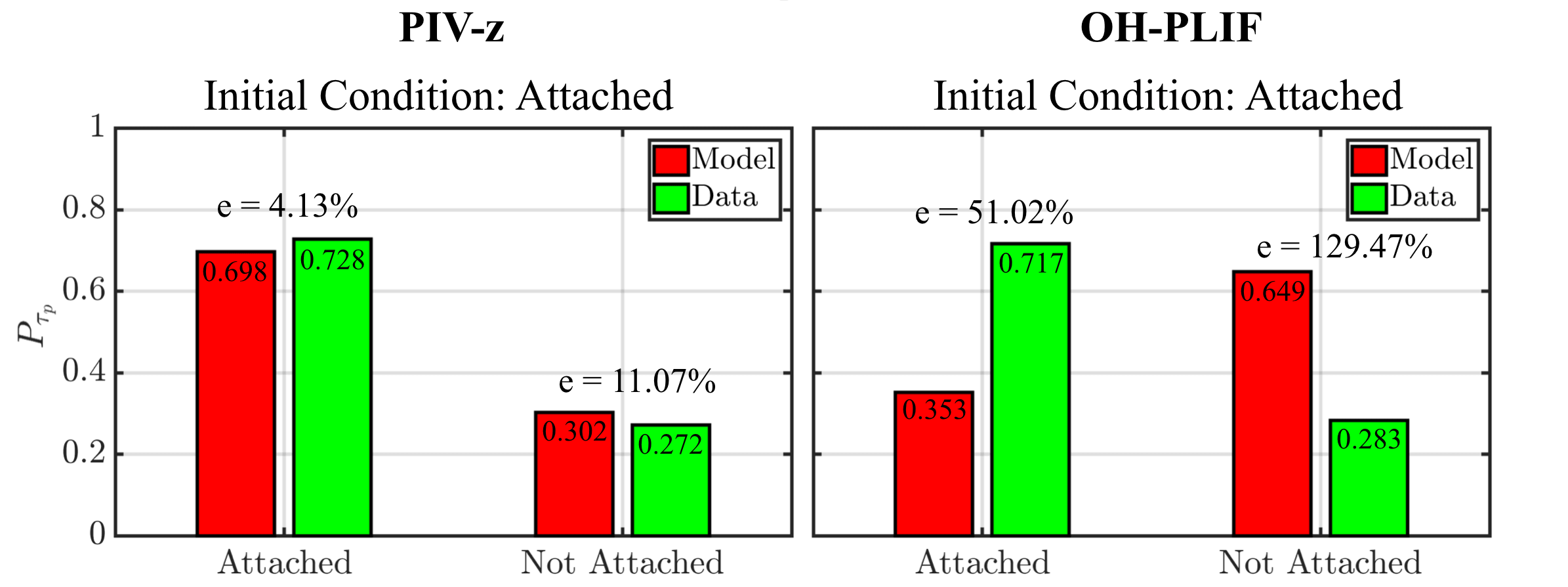}
      \caption{Forecast results for $\tau_p = \tau_f$ for out-of-plane velocity (left) and OH-PLIF (right) data, with the initial condition in the \textit{attached} state (flame detachment, or liftoff). Probability values indicated in bars, and relative precent error $e$ with respect to data-derived quantities is also shown. Y-axis is future probability at $\tau_p$. Results shown for $\tau_p/\tau_f=1$. }
      \label{fig:forecast_detachment}
\end{figure}

Detachment prediction results for $\tau_p = \tau_f$ are shown in Fig.~\ref{fig:forecast_detachment}. Note that final probabilities are in terms of two categories: 1) "attached" and 2) "not-attached". The category "attached" is essentially a persistence probability for remaining in the attached state after time $\tau_p$. The category "not-attached" includes probability of entering both transitioning and detached clusters from the attached state. As the quantity of interest in this case is the probability of \textit{exiting} the attached state, or transitioning out of the attached state, only these two categories are considered. In Fig.~\ref{fig:forecast_detachment}, plots on the left reflect PIV-z results, and tables on the right reflect OH-PLIF results. The results show predicted probabilities (red) and observed probabilities (green) for both data types. Note that, as alluded to above, the data-derived probabilities (green bars) across PIV-z and PLIF models are nearly equal. Relative percent errors $e$ between model and data-derived quantities are shown above the corresponding bars. Note that the PLIF model severely overshoots the probability of a snapshot leaving the attached state when compared to PIV-z (relative error of 129.47\% versus 11.07\%). It is clear that in forecasting flame detachment, PIV-z predictions are much more representative of the testing data than PLIF counterparts.  This is in line with previously observed experimental results showing OH-PLIF as a lagging indicator of PVC-induced strain rate, a direct catalyst for flame lift-off \cite{qiang}. 

\begin{figure}
      \centering
      \includegraphics[width=1\columnwidth]{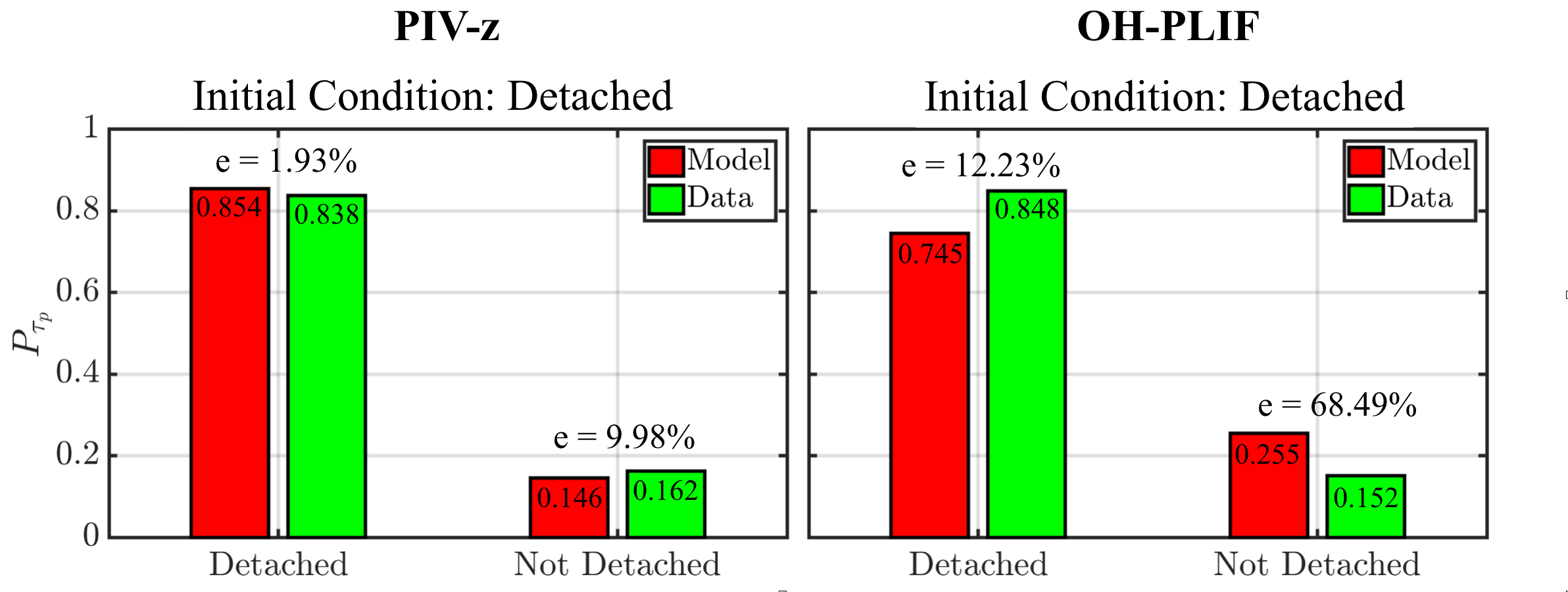}
      \caption{Forecast results for $\tau_p = \tau_f$ for out-of-plane velocity (left) and OH-PLIF (right) data, with the initial condition in the \textit{detached} state (flame attachment). Probability values indicated in bars, and relative percent error $e$ with respect to data-derived quantities is also shown. Y-axis is future probability at $\tau_p$. Results shown for $\tau_p/\tau_f=1$.}
      \label{fig:forecast_attachment}
\end{figure}

When predicting flame attachment for $\tau_p = \tau_f$ in Fig.~\ref{fig:forecast_detachment}, errors associated with the OH-PLIF models are again expectedly higher. Note that the data-derived probabilities also match to a reasonable degree across the different models. An interesting feature to note is the increased accuracy of models across the board for prediction of flame attachment (Fig.~\ref{fig:forecast_attachment}) versus flame detachment (Fig.~\ref{fig:forecast_detachment}): both PIV-z and PLIF model relative errors observe notable drops in relative error. Nevertheless, results show that in both detachment and attachment predictions, PIV-z produces more accurate forecasts when compared to the PLIF model.  

\begin{figure}
      \centering
      \includegraphics[width=1\columnwidth]{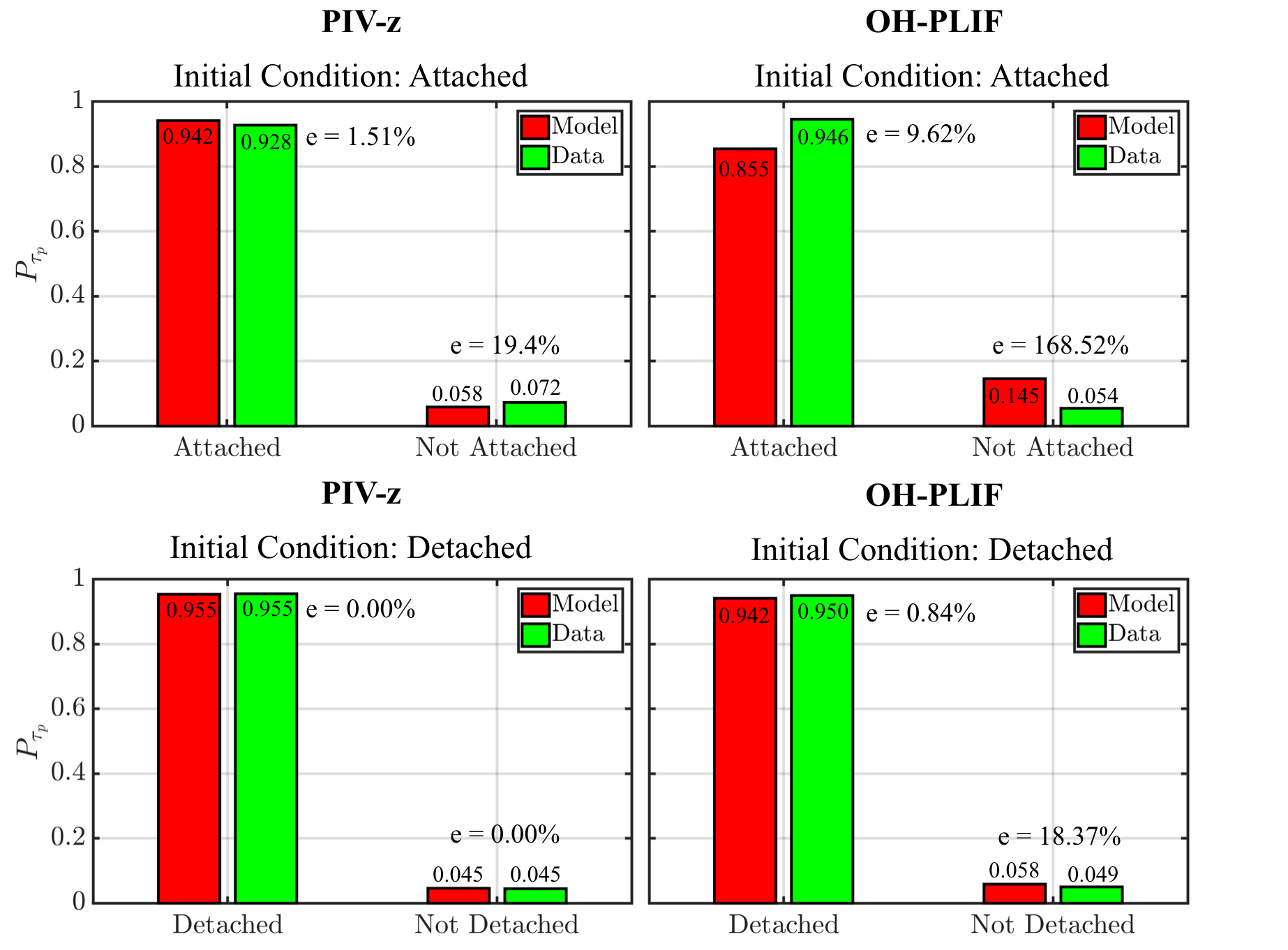}
      \caption{Short-time forecast results for $\tau_p = 0.1\tau_f$ for out-of-plane velocity (left) and OH-PLIF (right) data. Upper row is detachment forecast and lower row is attachment. Annotations made in same manner as Figs.~\ref{fig:forecast_shorttime} and \ref{fig:forecast_longtime}.}
      \label{fig:forecast_shorttime}
\end{figure}

\begin{figure}
      \centering
      \includegraphics[width=1\columnwidth]{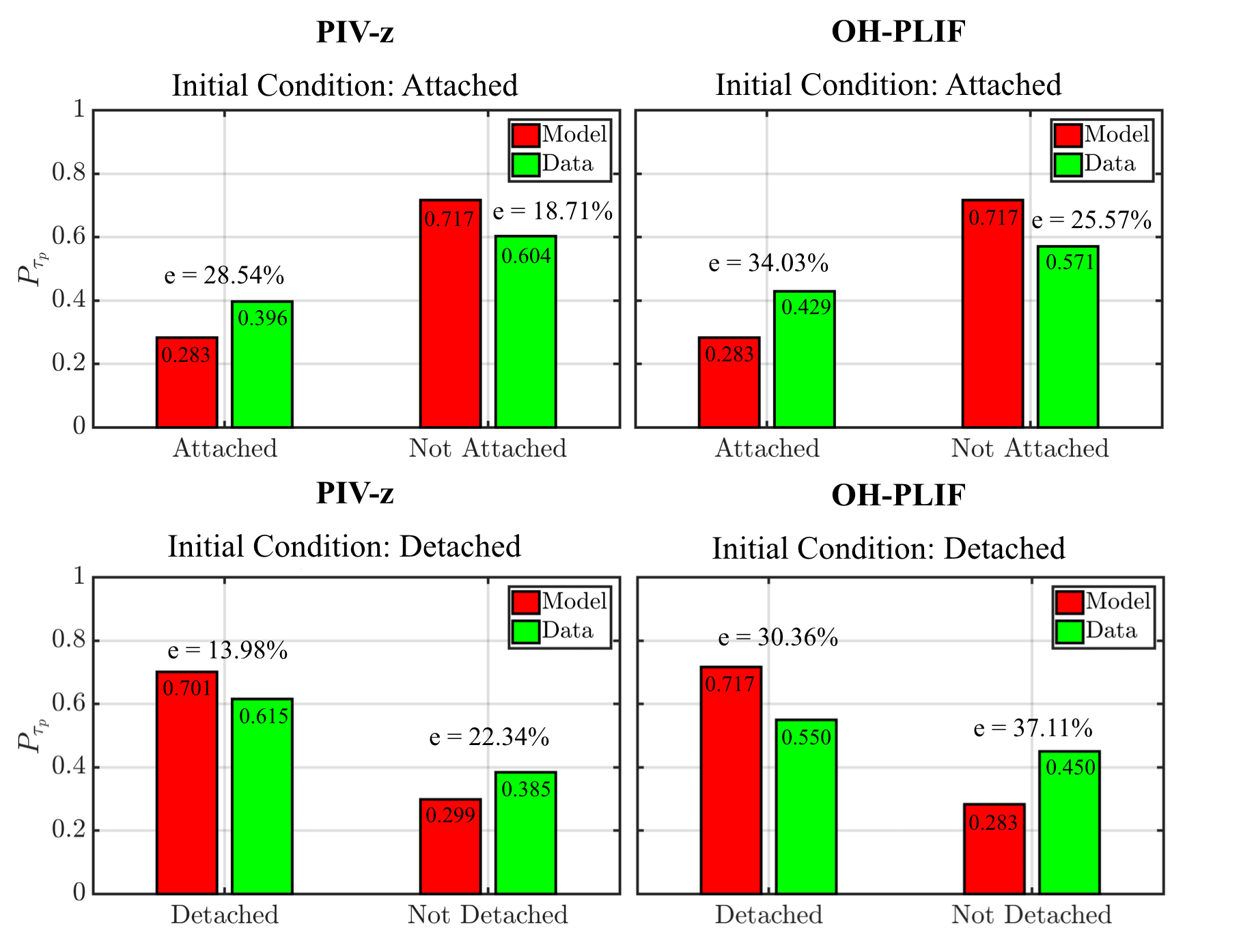}
      \caption{Long-time forecast results for $\tau_p = 7.0\tau_f$ for out-of-plane velocity (left) and OH-PLIF (right) data. Upper row is detachment forecast and lower row is attachment. Annotations made in same manner as Figs.~\ref{fig:forecast_shorttime} and \ref{fig:forecast_longtime}.}
      \label{fig:forecast_longtime}
\end{figure}

Prediction results for $\tau_p = 0.1\tau_f$ and $\tau_p = 7.0\tau_f$ are shown in Figs.~\ref{fig:forecast_shorttime} and \ref{fig:forecast_longtime}, respectively. In each of these figures, the flame detachment predictions (similar to Fig.~\ref{fig:forecast_detachment}) are shown in the upper row and attachment (similar to Fig.~\ref{fig:forecast_attachment}) in the lower row. As in the $\tau_p = \tau_f$ predictions, the short-time predictions in Fig.~\ref{fig:forecast_shorttime} show a similar trend of higher PIV-z model performance in the detachment process. Because the forecasting time is smaller by a factor of 10, this effect is not seen to the same level as in Fig.~\ref{fig:detachment_process}. Despite the slightly lower level of performance here for OH-PLIF, there is high accuracy in both models in short-time predictions -- they both forecast the expected state persistence with minimal relative errors. In fact, the predictions in Fig.~\ref{fig:forecast_shorttime} for PIV-z in the attachment process gives zero relative error. Note that probabilities for the "not-attached" and "not-detached" labels are low in Fig.~\ref{fig:forecast_shorttime} because the prediction time is extremely small. 

The long-time predictions in Fig.~\ref{fig:forecast_longtime} are provided to showcase CROM-based forecasting for timescales at or beyond model horizon times. Although trends of significantly lower OH-PLIF performance are seen here in the attachment process again, in the long-time prediction setting, both PIV-z and OH-PLIF model predictions show similar patterns. For example, they overshoot transition probability in the detachment process and undershoot in the attachment process. The model predictions (red bars) between PIV-z and PLIF are similar (especially in the detachment process) because the limit of the horizon time for both models has been reached. In other words, in the long-time limit, the models will always approximate the initial distribution of snapshots based on the training data (see Eq.~\ref{eq:q_vector}). Additionally, if the initial distribution does not perfectly match the ergodic distribution of the Markov chains, and if there are differences between the ergodic distributions in the training and testing data, the discrepancy in the CROM predictions at times at or beyond the prediction horizon is expected to rise even higher.

\begin{figure}
      \centering
      \includegraphics[width=0.6\columnwidth]{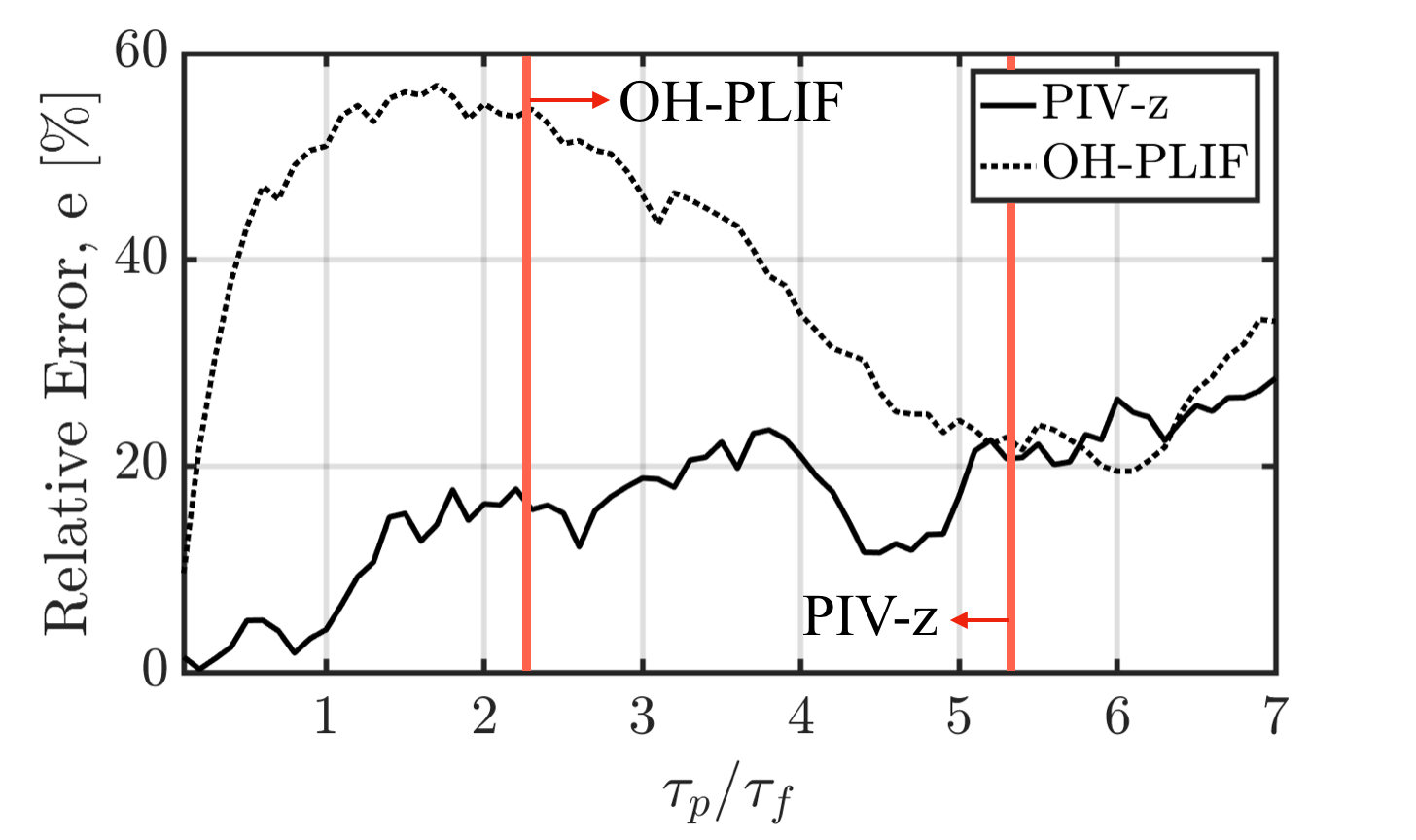}
      \caption{Relative error, $e$, as a function of normalized prediction time for the "attached" label in the detachment process. Vertical red lines indicate horizon times.}
      \label{fig:error_variation}
\end{figure}

For a clear illustration of prediction time effect on model performance, the variation in relative errors as a function of prediction time for the "attached" label in the detachment process is shown for PIV-z and OH-PLIF in Fig.~\ref{fig:error_variation}. The red vertical lines indicate the horizon times for the respective models. As evidenced in the discussion above, the plot shows large performance advantage in the PIV-z model in the ranges of $\tau_p / \tau_f = 0.1$ to about $\tau_p / \tau_f = 5.0$, after which both errors appear to converge. Note that this apparent convergence occurs before the PIV-z horizon time and after the OH-PLIF horizon time. This importantly verifies that the prediction horizon time can indeed be used as indicator of forecasting strength when considering prediction times smaller than the horizon times. An important result from Fig.~\ref{fig:error_variation} is that although one can use horizon time as a metric for forecasting strength, the horizon time itself is not a prediction time at which one should necessarily expect accurate forecasts. Note that the OH-PLIF curve increases until a peak near its horizon time, followed by a decrease until the convergence point. In contrast, the PIV-z curve appears to increase with heavy fluctuations. The intricacies of such behaviors require further study and are out-of-scope here, though this is an object of future work.

Ultimately, with the test conditions and model inputs used, out-of-plane velocity can be comfortably classified as the most potent dataset with regards to flame transition prognostics. This essentially hints that the prediction horizon time $\tau_h$ is a good indicator to quantify the predictive strength of dataset.

\section{Conclusions}
\label{sec:conclusion}

A cluster-based ROM was employed to analyze experimental data of flame transitions in a swirling combustor, and to create a model to anticipate such transitions. As an analysis tool, the method allows for the classification of data into modes that are statistically representative of the flow field and extraction of the most probable path between these modes during flame transitions. An appealing property of this process is that it can be done in an unsupervised manner, making the analysis as objective as possible. However, in order to use CROM, the number of modes must be set directly by the user. Two different approaches to set this number were presented in this work.

For the swirl combustor analyzed here, the flame detachment and attachment process were both analyzed by means of the transition matrix and modes outputted by CROM. It was shown how the modes could be associated to a macroscopic flame states (attached, detached, transition) and thereby derive a physical interpretation from the CROM output. It was found that the flame detachment process stems from an asymmetry in the flow field and for the flame. This suggests that an asymmetric process causes the flame detachment. This finding is in line with prior experimental analysis which identified PVC as the cause of flame detachment. The flame attachment appears to be triggered by a recirculation zone located at the same axial location as the flame, far from the nozzle. This recirculation seems to drive the flame down next to the nozzle. This analysis illustrates the potential of data-driven methods to analyze complex flows in a systematic way while highlighting key processes.

For forecasting purposes, the prediction horizon time of the transition matrix can be computed to help quantify the predictive power of a particular dataset. The predictive capabilities of the OH concentration and the three velocity components were compared. It was shown that z-component of velocity (swirling velocity) holds the largest prediction power while the OH signal leads to the shortest prediction window. This is consistent with the experimental observation that the cause of the flame transitions lies in the velocity field while the flame simply responds to strain rates. The present methodology allows to confirm that the swirling velocity is a better causal indicator than the OH signal. Using the cluster labeling method, the forecasting can be formulated in terms of the macroscopic flame state rather than individual modes. The PIV-z based model significantly increased prediction accuracy while the PLIF-based model misrepresents probability considerably. The prediction horizon time is therefore a good indicator for the predictive power of a dataset. Interestingly, both PIV-z and PLIF data accurately captured the probability of attachment process. This suggests that a different physical process involving OH concentration might govern the flame attachment.

This study introduces many pathways for future work. The horizon time analysis, as it was extended to analyze forecasting strength of specific data types, can also be used to search for an optimal sensor that would hold the best predictive power. Furthermore, the forecasting tool is an important step towards constructing actuation or other control strategies in combustors. In particular, this model can used to determine acceptable probability thresholds for initiating actuation procedure, and also to understand what part of the flow to inhibit in the actuation process. While the CROM application here shows considerable promise, there is a need to fully understand how the clustering techniques can be ``goal-oriented" in order to optimize the prediction for certain purposes only. The type of data is best suited for CROM forecasts was found by comparing OH-PLIF and PIV, but a similar study can also be conducted by comparing different spatial locations, proceeding with an analysis of the effect of prediction horizon time on model accuracy. These topics will be considered in future work.

\addcontentsline{toc}{section}{References}
\bibliographystyle{plain}
\bibliography{main.bbl}

\appendix
\section{Model Validation}
\label{sec:markov_assumption}

There are underlying assumptions in the predictive model obtained with the CROM method. First, in order to use the model to predict unseen data, the model should have been trained on a dataset that is representative of most of the combustor dynamics. Second, the transition from one time-step to the next is assumed to be Markovian. Here, the validity of these assumptions is assessed. To this end, two tests are conducted and applied to the PIV-z dataset only for conciseness. The same trends hold for all other datasets. 

\begin{figure}
    \centering
    \includegraphics[width=0.7\columnwidth]{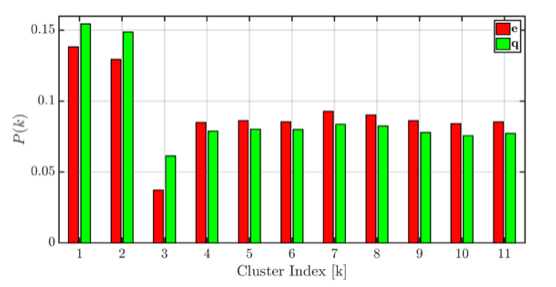}
    \caption{Test 1: comparison between ergodic distribution from model ($\bm e$) and initial cluster distribution from data ($\bm q$).}
    \label{fig:pivz_distribution}
\end{figure}

In the first test, the stationary PDF obtained from the transition matrix $\bm e$ (eigenvector of the unit eigenvalue) is compared to the PDF of clusters obtained from the dataset $\bm q$~\cite{schneider_markov}. In order to have good agreement between both PDFs, the interplay between the fast and slow timescales of the combustor need to be accurately represented by the transition matrix. The acceptable agreement shown by Fig.~\ref{fig:pivz_distribution} suggests that the training data was sufficient to capture most of the relevant dynamics and allow to accurately describe the long term dynamics of the swirler. 

\begin{figure}
    \centering
    \includegraphics[width=1\columnwidth]{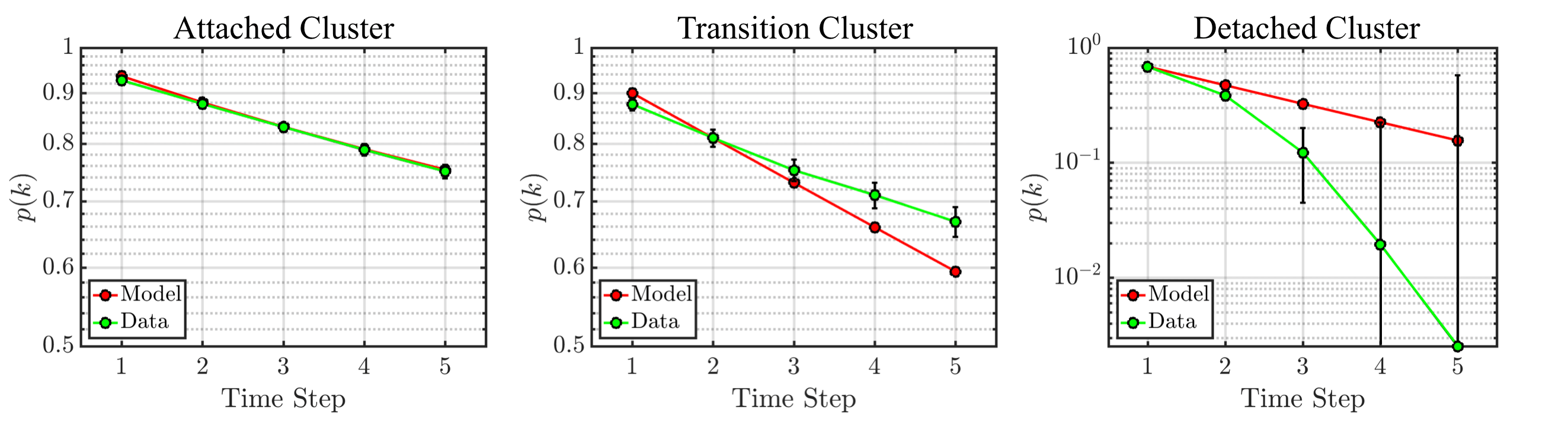}
    \caption{Test 2: persistence time plot from the PIV-z dataset, where y-axis is probability to remain in the same cluster and x-axis is time-step.}
    \label{fig:pivz_persistence}
\end{figure}

The second test consists in comparing short-term forecasts obtained from the transition matrix and the data used to generate the model. Similar to what was done in Ref.~\cite{schneider_markov}, the persistence probability of the clusters is compared between the model and the data. The persistence probability is defined as the probability for a snapshot from some cluster to remain in this cluster after a certain number of time-steps. If the persistence probability given by the model does not deviate from that given by the data by a significant amount, the Markov assumption is valid. It is reminded to the reader that the training data is defined for $t = [0.0, 1.0]$ seconds (10000 snapshots) and the testing data is taken to be $t = [1.0, 1.5]$ seconds (5000 snapshots). Fig.~\ref{fig:pivz_persistence} shows the persistence probability for three clusters (one for each flame state) plotted against time. Note that statistical uncertainties are included for probabilities extracted from data, with growing uncertainties for shrinking probabilities. Interestingly, model-data agreement is nearly perfect in persistence probability for an attached cluster, and is fairly well-captured for a transition cluster. In the detached case, data-based probabilities diverge quickly from the model counterparts, though uncertainties drastically increase as time-step increases. The severe drop in probability observed by the data-derived curve (in green) in the detached cluster is expected, as a snapshot starting in a detached cluster has a high chance to move on to the next detached cluster in the data, leading to very small persistence probability with increasing time-step. Refining the detached state with a higher $N_k$ would increase this effect. The model is arguably accurate for certain cases but has imperfections for others. It is unclear if these are due to the physics involved or the statistical uncertainties of the testing data for the low persistence probabilities of the detached clusters. The imperfect agreement here, though still acceptable, should be kept in mind when interpreting the actual transition forecasts made in Sec.~\ref{sec:results-forecasting}. 

\end{document}